\documentstyle[epsfig]{elsart}
\begin{document}

\newcommand{\re}{\mathop{\mathrm{Re}}}
\newcommand{\im}{\mathop{\mathrm{Im}}}
\newcommand{\D}{\mathop{\mathrm{d}}}
\newcommand{\I}{\mathop{\mathrm{i}}}


DESY 03-108

August 2003
\bigskip

\begin{frontmatter}

\journal{}
\date{}

\title{
The Free Electron Laser Klystron Amplifier Concept}

\author[DESY]{E.L.~Saldin},
\author[DESY]{E.A.~Schneidmiller},
and \author[Dubna]{M.V.~Yurkov}

\address[DESY]{Deutsches Elektronen-Synchrotron (DESY),
Notkestrasse 85, D-22607 Hamburg, Germany}

\address[Dubna]{Joint Institute for Nuclear Research, Dubna,
141980 Moscow Region, Russia}

\begin{abstract}

The simplest high gain free electron laser (FEL)
amplifier concept is proposed.  A klystron amplifier has the useful
property that the various electronic processes take place in separate
portions of the amplifier, rather than overlapping as in  FEL amplifier
with an uniform undulator.  The klystron consists of two fundamental
parts: succession of 2-3 cascades (modulator), and
an output undulator (radiator) in which the modulated electron beam
coherently radiates. Each cascade consists of uniform undulator and
dispersion section. Unlike distributed optical klystrons, we have a
high gain per cascade pass.  This has a few consequences.
First, klystron gain does not depend on the bunch compression
in the injector linac, i.e. maximum gain per cascade pass at high peak
beam current is the same as at low peak beam current, without
compression.  Conventional, short-wavelength FEL amplifier and
distributed optical klystron require electron beam peak current of a
few kA. Second, there are no requirements on the alignment of the
cascade undulators and dispersion sections, because, in our (high gain)
case, there is no need for radiation phase matching. There are
applications, like XFELs, where the unique properties of high gain
klystron FEL amplifier are very desirable.  Such a scheme allows one to
decrease the total length of the magnetic system. In this paper we
discuss also implementation of the proposed SASE FEL scheme for the
frequency multiplier, for femtosecond experiments, and multi-user soft
X-ray facility.

\end{abstract}

\end{frontmatter}

\clearpage

\section{Introduction}

\begin{figure}[tb]
\begin{center}
\epsfig{file=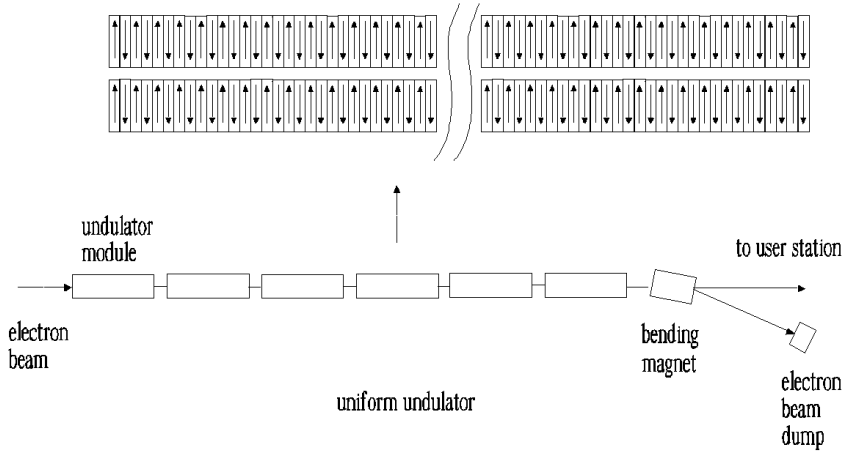,width=0.8\textwidth}
\end{center}
\caption{Schematic diagram of conventional SASE FEL
}
\label{fig:a1}
\end{figure}

\begin{figure}[tb]
\begin{center}
\epsfig{file=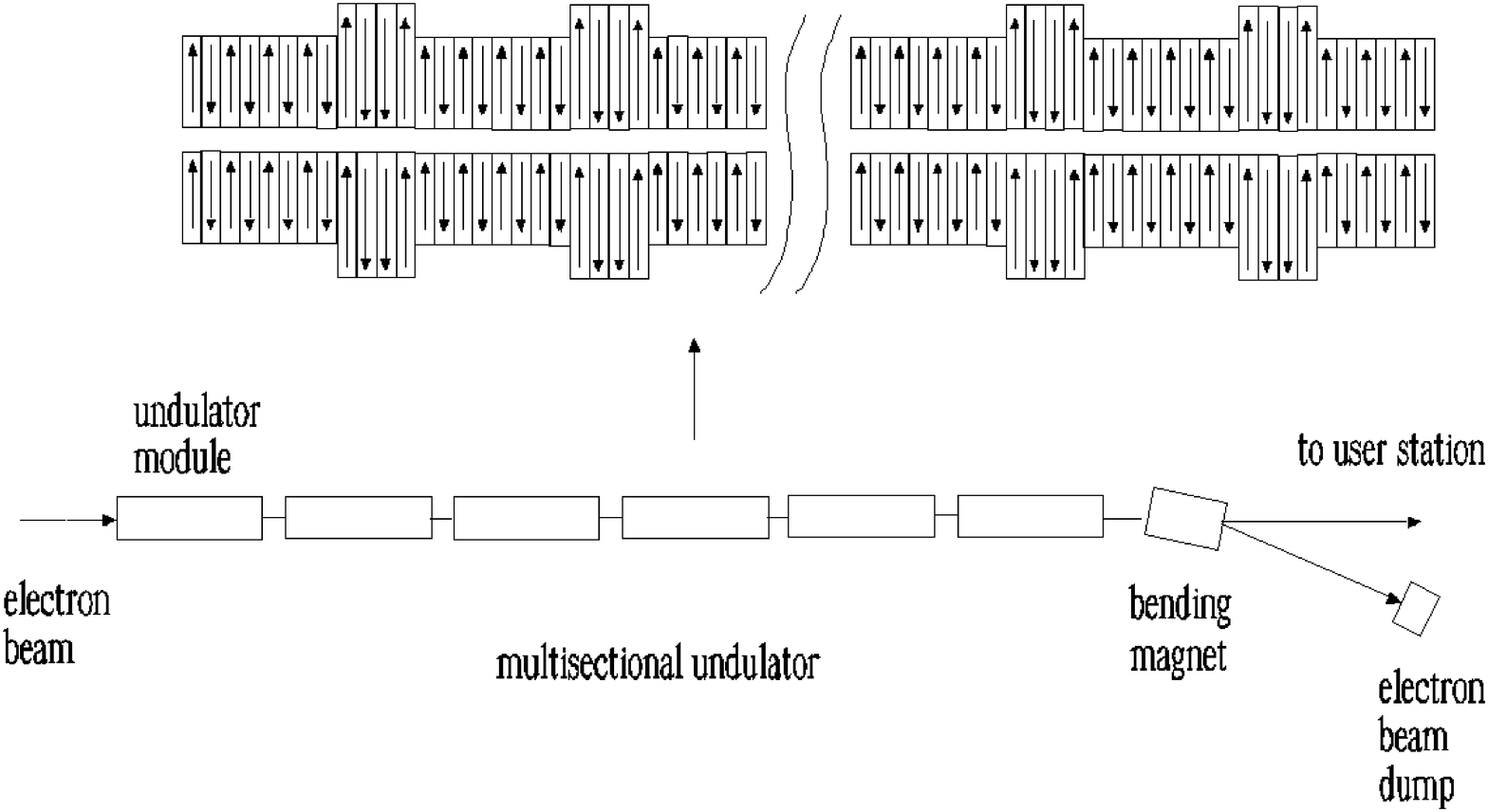,width=0.8\textwidth}
\end{center}
\caption{Schematic diagram of distributed optical klystron
}
\label{fig:a2}
\end{figure}

\begin{figure}[tb]
\begin{center}
\epsfig{file=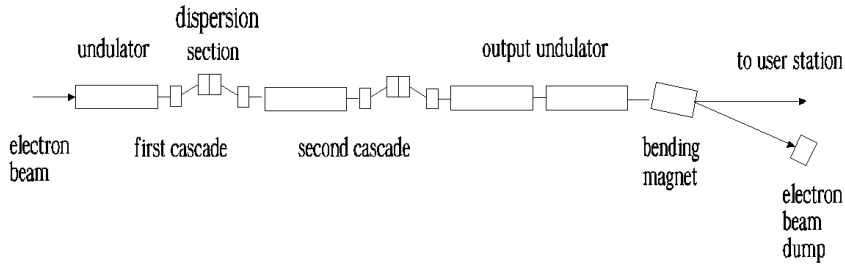,width=0.8\textwidth}
\end{center}
\caption{Schematic diagram of high gain cascade klystron amplifier
}
\label{fig:a}
\end{figure}

High gain FEL amplifiers are of interest for a variety of potential
applications that range from X-ray lasers \cite{x1,x2} to ultraviolet
MW-scale industrial lasers \cite{m}. There are various versions of the
high gain FEL amplifier. A number of high gain FEL amplifier concepts
may prove useful for XFEL applications.  Two especially noteworthy ones
are the FEL amplifier with a single uniform undulator \cite{x1,x2} (see
Fig.  \ref{fig:a1}) and the distributed optical klystron \cite{p,l,b,n}
(see Fig. \ref{fig:a2}).  The high gain cascade klystron amplifier
described in this paper is an attractive alternative to other
configurations for operation in the X-ray wavelength range (see Fig.
\ref{fig:a}).

Let us first investigate qualitatively the processes that take place in
any FEL amplifier; this investigation should also enable us to
distinguish somewhat more accurately between the so-called
distributed optical klystron concept and the new high gain klystron
amplifier concept.  In any FEL, several processes must take
place, either separately or simultaneously. The first process that must
occur in any FEL is energy modulation. Energy modulation means
the application of radiation fields to the electrons as they move in
their trajectories under the action of undulator magnetic fields, in
such a way that the energies of these electrons are varied in some
periodic fashion. It is then necessary to take advantage of this
phase-dependent motion.

The second process, then, that must occure in a FEL is the conversion
of this variation in the motion of the electrons into a usable form. It
is necessary to turn the variations in the electron energy into a
radiation frequency current, which can deliver energy to the
electromagnetic field; i.e., the variations in velocity must be
converted into density variations. The ways in which this conversion is
brought about are numerous, and FEL
amplifier types are distinguished by their conversion
mechanism.

It may be useful to illustrate processes at this point to see how
energy modulation and conversion into density modulation take place
in some of the types of FEL amplifier.
In a conventional high gain FEL amplifier with a single uniform
undulator energy modulation and bunching occurs both simultaneously and
continuously along the undulator.  Energy modulation occurring at any
one point continues to affect bunching at points beyond, with
additional modulation being added as the beam moves along.
The problem of electromagnetic wave amplification in the uniform
undulator refers to a class of self-consistent problem. To solve the
problem, the field equations and equations of particle motion should be
solved simultaneously. When the FEL amplifier operates in
linear regime the transverse distribution of the radiation field
remains fixed, while its amplitude grows exponentially with undulator
length.

Perhaps the simplest FEL amplifier, from pedagogical point of view, is
the high gain klystron amplifier. This simplicity arises from the fact
that the important processes of energy modulation of the beam and
subsequent conversion of the energy modulation to density modulation
occur in distinctly separate regions of the FEL and thus may be
considered  in a sequential manner.  This separability of functions
aids materially in developing an understanding of modulation and
bunching phenomena.

In its simplest configuration the klystron amplifier consists of an
input undulator (energy modulator),
and an output undulator (radiator) separated by a dispersion section.
In the input undulator an incident
electron beam is subjected to a radiation field, which produces an
energy modulation, a modulation which after dispersion section, is
converted into an density modulation in the beam. In the output
undulator the bunched beam radiates coherently. This process, of
conversion of energy modulation to bunching by the action of a
dispersion section, is one of the characteristics of the klystron and
is one of the reasons why this device is so useful in many connections.

We have considered the two-undulator
klystron amplifier. In some experimental situations this simplest
configuration is not optimal. For example, this study has shown that in
the soft X-ray wavelength range the maximal (intensity) gain of this
amplifier does not exceed 30-40 dB. The obvious
and technically possible solution of the problem of gain increase might
be to use a klystron with three or more undulators.  Suppose that we
have a three undulator (or so-called two-cascade) klystron, and that
we introduce a very small signal into the first undulator.  This will
result in a small value of the amplitude of energy modulation, and a
resulting rather small power output from the second undulator.
Nevertheless, the radiation field amplitude in the second undulator
will still be much greater than that in the first undulator, on account
of the considerable amplification of the first cascade. We now
disregard the power produced by the second undulator, but consider its
effect on the electron beam passing through its magnetic system.  The
considerable radiation field in the second undulator will produce a
further energy modulation of the beam and, when it emerges on the far
side of the second undulator, it will have a large energy modulation,
which, with a second dispersion section, will result in optimum
bunching. This bunched beam then enters the third (radiator) undulator,
and produces a large amount of power.  In other words, we have
essentially combined two stages of amplification by incorporating
essentially two klystrons in a single unit.  The final power
is no greater than could be produced by a two-undulator klystron with
the same output undulator, but we can secure this power with a much
weaker input signal.

Let us consider now the problem of a
periodically spaced sequence of undulators and dispersion sections,
with the electron beam traversing each in succession. The larger the
number of cascades the longer is the amplifier length. Suppose that the
increase of total length (gain) of amplifier is tempered by an
accompanying decrease in the length (gain) of cascade.  From this
approach at very large number of cascades and small gain per cascade
pass we are led to the other device, which was the subject of the
papers \cite{p,l,b,n}:  the "distributed optical klystron".  At first
glance this is essentially the extension klystron amplifier, extended
to a very large number of cascades.  However, as noted
earlier, this amplifier is considerably more complicated in principle
than a klystron amplifier with high gain per
cascade pass.  When the gain per cascade pass is small, the cascades
cannot be considered independent.  It is important to realize that the
cascades are connected optically, so that the radiation produced, for
example, into the first undulator passes on into the second and
produces additional energy modulation of the beam. It is clear that we
cannot conveniently treat the theory of distributed klystron by the
same method we used in the klystron case.
Because of the low gain per cascade pass used for the distributed
klystron, it cannot be approximated as scheme with separate functions,
and the klystron amplifier theory has all based on that assumption. A
quite different approach to the problem, which proves to be more suited
to mathematical development of distributed klystron theory, is more
along the lines of an analogy with the theory of conventional FEL
amplifier with uniform undulator. The problem of electromagnetic wave
amplification in the distributed klystron refers to a class of
self-consistent problems. For instance, when the distributed klystron
operates in linear regime its amplitude gain grows exponentially with
cascade numbers.  From a study of the rate of exponential build-up of
the electromagnetic wave we can find the gain of the amplifier.

Our theory of klystron amplifier is based on the assumption that beam
density modulation does not appreciably change as the beam propagates
through the undulator. This approximation means that only the
contributions to the radiation field arising from the initial density
modulation are taken into account, and not those from the induced
bunching. At very long klystron undulators we are led to device,
which is the subject of the papers \cite{nv,ch}.
Two-segment high gain FEL amplifier is variant of the standard
high-gain FEL amplifier with uniform undulator. Because of the high
gain per undulator pass used for this modification, it cannot be
approximated as scheme with separate functions.

The klystron amplifier analysis is not only relevant from a
pedagogical point of view. There are reasons to expect that the
new concept of FEL amplifier discussed here can be of interest for
short wavelength applications such as X-ray SASE FELs.
Generation of X-ray radiation using klystron amplifiers has
many advantages, primarily because of the separation of electronic
processes, gain, and
potentially low sensitivity to alignment.  In principle, the gain of
the klystron amplifier can be increased indefinitely by adding
additional cascades along the beam.  In practice, with three-cascade
klystron amplifier, a gain in excess of 80 dB may be obtained in the
soft X-ray wavelength range.  At this level, the shot noise of the beam
is amplified up to saturation and a cascade klystron amplifier operates
as soft X-ray SASE FEL.

Now our intention is to develop certain physical ideas about the
operation of the klystron amplifier, the characteristics of its
radiation, and those features that set it apart from other FEL sources
in the short wavelength region of the spectrum. The resonance
properties of the radiation of the electron beam in the klystron can
be understood as follows. Since a dispersion section is a nonresonant
device, the klystron bandwidth is restricted by the resonance
properties of the cascade undulator, that is, by its number of periods
$N_{\mathrm{w}}$.  The typical amplification bandwidth of the klystron
in the soft X-ray wavelength range is on the order of one percent
$(N_{\mathrm{w}} \simeq 10^{2})$.  Since we study the start-up from
shot noise, we assume the input current to have a homogeneous spectral
distribution.  At large number of undulator periods the spectrum of
transversely coherent fraction of radiation is concentrated within the
narrow band, $\Delta\lambda/\lambda \simeq N^{-1}_{\mathrm{w}}$.

The actual physical picture of start-up from noise should take into
account that the fluctuations of current density in the electron beam
are uncorrelated not only in time but in space too. Thus a large number
of transverse radiation modes are excited when the electron beam enters
the undulator. These radiation modes have different gain.
Obviously, as cascade number progresses, the high gain modes will
predominate  more and more and we can regard the klystron as a filter,
in the sense that it filters from arbitrary radiation field those
components corresponding to the high gain modes. If we consider the
undulator radiation from the point of view of paraxial optics, then we
immediately see that the high gain modes are associated with radiation
propagating along the axis of the undulator, as opposed to radiation
propagating at an angle to the axis which has a low gain.
Consider an electron beam of radius
$r_{\mathrm{b}}$ moving in the (cascade) undulator with length
$L_{\mathrm{w}}$.  The parameter
$N_{\mathrm{F}} = r_{\mathrm{b}}^{2}/(\lambda L_{\mathrm{w}})$ can be
referred to as the Fresnel number. Here $\lambda$ is the radiation
wavelength. The requirement for the mode selection to be quickly holds
at small value of the Fresnel number.  Hence, for thin electron beam
with Fresnel number of $N_{\mathrm{F}} \simeq 1$, the emission will
emerge in a single (fundamental) transverse mode and the degree of
transverse coherence of output radiation will approach unity.

As mentioned above, amplification in each cascade is
a two-step process. Since we study the
start-up from shot noise there is no electromagnetic field at the
undulator entrance and the modulation of the beam density serves as the
input signal for the klystron amplifier.
The radiation field produced an
ponderomotive potential across the undulator; as is now known from
general FEL theory, this voltage modulates the energy of the electron
beam. Then the energy modulation causes modulation of the beam density
in the dispersion section.  If we can neglect beam bunching in the
cascade undulator, the amplitude of energy modulation at the undulator
exit, $\delta{\cal E}$, is proportional to the resonance Fourier
component of beam current, $\delta I_{\mathrm{in}} =
a_{\mathrm{in}}I_{0}$.  Here $a_{\mathrm{in}}$ is the initial beam
microbunching at the resonant wavelength, and
$I_{0}$ is the peak  beam current.  We now examine the behaviour of the
electrons as they travel in a  dispersion section.  Like the situation
with usual microwave klystron, the Bessel function factor $J_{1}(X)$
represents the microbunching at the dispersion section exit.  The
parameter $X$ in our case is proportional to the product of the
amplitude of energy modulation $\delta{\cal E}$ and the compaction
factor of the dispersion section $R_{56}$. In the high gain linear
regime $X$ is much smaller then unity and  we
make the approximation $J_{1}(X) \simeq X/2$. Thus, microbunching at
the cascade exit is proportional to input signal amplitude,
$a_{\mathrm{out}} \sim R_{56} I_{0} a_{\mathrm{in}}$. When a cascade
operates in the high gain linear regime the gain per cascade pass is
independent of input signal amplitude and proportional to the product
of the compaction factor and the peak beam current, $G =
a_{\mathrm{out}}/a_{\mathrm{in}} \sim R_{56} I_{0}$.

If we build a klystron, we want to have
small number of cascades and high gain.  One might think that all we
have to do is to get enough gain per cascade pass - we can always
increase compaction factor $R_{56}$ and we can always increase $G$, and
there is no reason why we cannot use the simple two-undulator klystron
scheme.  So the limitations of a gain per cascade pass are not that it
is impossible to build a dispersion section that has a very large
compaction factor.   In fact, one of the main problems in klystron
amplifier operation is preventing spread of microbunching due to energy
spread into the electron beam. This energy spread is unimportant when
the compaction factor of dispersion section is relatively small, but we
shell see that energy spread  is of fundamental importance in the case
of very large compaction factor.
To get a rough idea of the spread of the microbunching, the position of
the particles within the electron bunch at the dispersion section exit
has a spread which is equal to $\Delta z \simeq R_{56}\Delta\gamma
/\gamma$, where $\Delta\gamma/\gamma$ is the local energy spread into
the electron beam. And how big is $\Delta z$ in the spread-out
microbunching? We know that uncertainty in the phase of the particles
is about $\Delta\psi \simeq 2\pi\Delta z/\lambda$. Therefore, the
largest compaction factor that we can have is approximately $R_{56}
\simeq \lambda\gamma/(2\pi\Delta\gamma )$. This is what we wanted to
deduce - that the gain maximum per cascade pass varies as ratio
$G_{\mathrm{max}} \sim I_{0}/\Delta\gamma$.  It is of course
desirable that the klystron gain be maximum.  An experimenter can
easily tune the gain of the klystron by tuning the magnetic fields in
the dispersion sections.  It should be pointed out that the local
energy spread plays a very important role in the operation of the FEL
klystron amplifier, and this characteristic distinguishes the klystron
from conventional FEL amplifier.

Electron bunches with very small transverse emittance and high peak
current are needed for the operation of conventional XFELs. This is
achieved using a two-step strategy: first generate beams with small
transverse emittance using an RF photocathode and, second, apply
longitudinal compression at high energy using a magnetic chicane.
Although simple in first-order theory, the physics of bunch compression
becomes very challenging if collective effects like space charge forces
and coherent synchrotron radiation forces (CSR) are taken into account.
Wakefields of bunches as short as 0.1 mm have never been measured and
are challenging to predict.

The situation is quite different for klystron amplifier scheme
described in our paper. An distinguishing feature of the klystron
amplifier is the absence of apparent limitations which would
prevent operation without bunch compression in the injector linac.
According to our discussion above, the gain maximum
per cascade pass is proportional to the peak current and inversely
proportional to the energy spread of the beam.  Since the bunch length
and energy spread are related to each other through Liouville's
theorem, the peak current and energy spread cannot vary independently
of each other in the injector linac.  To extent that local energy spead
is proportional to the peak current, which is usually the case for
bunch compression, the gain will be independent of the actual peak
current.  We see, therefore, that klystron gain in linear regime
depends only on the actual photoinjector parameters.
This incipient proportionality between gain and
$I_{0}/\Delta\gamma$ is a temptation, in designing an XFEL, to go to
very high values of $I_{0}/\Delta\gamma$ and very long values of bunch
length.

Let us present a specific numerical example for the case of a
4th generation soft X-ray light source based on the use RF
photoinjector and superconducting linac with duty factor 1 \%.
The average brilliance of a klystron facility
operating without bunch compression in the injector linac surpasses the
spontaneous undulator radiation from 3rd generation synchrotron
radiation facilities by 4 or more orders of magnitude.
Decreasing the peak current also decreases the peak brilliance of
the SASE FEL radiation by about of factor 100, but this is still 6
order of magnitude higher than that of 3rd generation synchrotron
radiation sources.

Our studies have shown that the soft X-ray cascade
klystron holds great promise as a source of radiation for generating
high power single femtosecond pulses. The obvious temporal limitation
of the visible pump/X-ray probe technique is the duration of the X-ray
probe.  At a klystron facility operating without bunch compression, the
X-ray pulse duration is about 10 ps.  This is longer than the timescale
of many interesting physical phenomena. The new principle
of pump-probe techniques described in section 6 offers a way around
this difficulty. Section 6 also deals with the design strategy for the
multi-user distribution system for an X-ray laboratory. An X-ray
laboratory should serve several, may be up to ten experimental
stations which can be operated independently according to the needs of
the user community. On the other hand, the prefered layout of a
conventional SASE FEL is a linear arrangement in which the injector ,
accelerator, bunch compressors and undulators are nearly collinear, and
in which the electron beam does not change direction between
accelerator and undulators. The situation is quite different for
the klystron amplifier scheme proposed in our paper.  Since it operates
without bunch compression in the injector linac, the problem of
emittance dilution in the bending magnets does not exist. An
electron beam distribution system based on unbunched electron beam can
provide efficient ways to generate a multi-user facility - very similar
to present day synchrotron radiation facilities.

It may be wondered, after hearing of all of the wonderful properties of
the high gain klystron amplifier, why this  advance in XFEL
techniques occurred only this year. It should be note that the XFEL
which was proposed in 1982 \cite{first} has had nearly 20 years of
development. One of the explanations is as follows. For years we were
led to believe that local energy spread into electron beam which
delivered from a photoinjector is relatively large. Recent analysis of
experimental results obtained at TTF SASE FEL \cite{a} shows that the
value of the local energy spread should be revised.  After bunch
compression, it is expected to be about 0.2 MeV (for $I_{0} \simeq 2.5
\ {\mathrm{kA}}$) or smaller which is significantly less than the
previously projected value of 1 MeV. This
decrease significantly improves operation of klystron amplifier and
extends the safety margin for  X-ray klystron facility operation.

This paper is
intended primarily for the experimental
scientist or XFEL program manager who has to design XFELs and who
wishes to undestand the operating principles of XFELs. The material to
be discussed provides the necessary background information required for
making strategic XFEL design decisions. In this paper the electronic
processes that occur in an XFEL will be discussed in elementary fashion.
The XFEL theory has a, not
entirely undeserved, reputation for difficulty and obscurity. This is
due partly to the complexity of systems in which (shot) noise is input
signal and partly to the unfamiliar nature of some of the mathematical
tools involved (integrodifferential equations). Klystron amplifiers
are one of the simplest examples of high gain FEL amplifiers and
therefore we shall use them as an introduction to the techniques of
XFEL theory.  In the klystron amplifier case little demand is made
on the reader's mathematical ability.  The mathematics that is involved
is particularly simple, involving very simple differential equations
and algebraic operations and no integrodifferential equations. Such
introduction to the operating principles of XFELs has never been done
before, to our knowledge.

\section{Interaction between electrons and fields in an undulator}

Klystron amplifiers amplify the input signal. To describe the
klystron amplifier operation, we first define the initial conditions.
Important practical kind of initial condition refers to the case when
there is no radiation field at the undulator entrance and the
modulation of the beam current density serves as the input signal for
the amplifier. As stated in the introductory section, in undulator two
physical processes take place simultaneously:

\noindent 1) The modulated electron beam excites electromagnetic waves
which are propagated along the undulator;

\noindent 2) The electric field components of these waves produce a
modulation of the energy of the electrons.

\noindent The approach discussed in this section consists of deriving
for these two process separate expressions which, when joined, give
quantitative information about the composite phenomena taking place in
the undulator. We assume that the density modulation effect is
negligible. Hence, electrons passing through the first undulator
receive only energy modulation, but no density modulation. Density
modulation present when the electrons enter the second undulator, must
have developed while the electrons were traveling in the dispersion
section.  This effect will be discussed later.

\subsection{Collective fields produced by the modulated electron beam
in an undulator}

The simplest form of bunched beam is found if we have swarm of
electrons, all moving in the same direction (say the $z$-direction)
with the same velocity (say $v$), but in which the number of
electrons per unit volume (say $n$) depends on $z$. Then if the charge
on the electron $(-e)$ we can easily find the current density at any
value of $z$. The quantity $n$ must really be a function of
$z/v-t$, on account of the velocity $v$ of the electrons. Thus,
the current density at point $z$, at time $t$, is
$-ev_{z}n(z/v-t)$, where $n(z/v-t)$ indicates the function $n$
of the argument $z/v-t$.  Unless $n$ is a constant independent of
its argument, this will lead to a current density which varies with
time.  The simplest variation of $n$ for our present purposes is a
superposition of a constant, or d-c component, and a periodic function
of $z/v-t$, with an angular frequency $\omega$.  First we consider
the case where $n$ is simply a constant plus a single cosine term:
$n(z/v-t) = n_{0}[1 + a_{\mathrm{in}}\cos\omega(z/v-t)]$.
We must note one important fact: $n$ can never
become negative, and hence $a_{\mathrm{in}}$ cannot be greater than
unity.

To understand the basic principles of klystron amplifier operation, let
us consider the helical undulator. The magnetic field on the axis of
the helical undulator is given by

\begin{displaymath}
\vec{H}_{\mathrm{w}} = \vec{e}_{x}H_{\mathrm{w}}\cos(k_{\mathrm{w}}z) -
\vec{e}_{y}H_{\mathrm{w}}\sin(k_{\mathrm{w}}z) \ ,
\end{displaymath}

\noindent where $k_{\mathrm{w}} = 2\pi/\lambda_{\mathrm{w}}$ is the
undulator wavenumber and $\vec{e}_{x,y}$ are unit vectors directed
along the $x$ and $y$ axes of the Cartesian coordinate system
$(x,y,z)$. The Lorentz force $\vec{F} = -
e(\vec{v}\times\vec{H}_{\mathrm{w}})/c$ is used to derive the
equations of motion of electrons with charge $(-e)$ and mass
$m_{\mathrm{e}}$ in the presence of the magnetic field.
The explicit expression for the electron velocity in the
field of the helical undulator has the form:

\begin{displaymath}
\vec{v}_{\perp}(z) = c\theta_{\mathrm{w}}\left[
\vec{e}_{x}\cos(k_{\mathrm{w}}z) -
\vec{e}_{y}\sin(k_{\mathrm{w}}z)\right] \ ,
\end{displaymath}

\noindent which means that the electron in the undulator moves along
a constrained helical trajectory parallel to the $z$ axis. The angle
of rotation is given by the relation
$\theta_{\mathrm{w}} = K/\gamma =
\lambda_{\mathrm{w}}eH_{\mathrm{w}}/(2\pi
m_{\mathrm{e}}c^{2}\gamma)$, where $\gamma = (1-v^{2}/c^{2})^{-1/2}$ is
the relativistic factor and $v^{2} = v_{x}^{2} + v_{y}^{2} +
v_{z}^{2}$. As a rule, the electron rotation angle
$\theta_{\mathrm{w}}$ is small and the longitudinal electron velocity
$v_{z}$ is close to the velocity of light, $v_{z} \simeq c$.

Let us consider a bunched relativistic electron beam moving along the
$z$ axis in the field of a helical undulator. In what follows we use
the following assumptions:  i) the electrons move along constrained
helical trajectories in parallel with the $z$ axis; ii) the radius of
the electron rotation in the undulator, $r_{\mathrm{w}} =
\theta_{\mathrm{w}}/k_{\mathrm{w}}$, is much less than the transverse
size of the electron beam. Next let us assume that electron beam
density at the undulator entrance is simply  $n =
n_{0}(\vec{r}_{\perp})[1 + a_{\mathrm{in}}\cos\omega(z/v_{z}-t)]$,
where $a_{\mathrm{in}} = {\mathrm{const.}}$ In other words we consider
the case in which there are no variations in amplitude and phase of
the density modulation in the transverse plane.

Under these assumptions the transverse
current density may be written in the form

\begin{displaymath}
\vec{j}_{\perp} = - e\vec{v}_{\perp}(z)n(\vec{r}_{\perp}, z/v_{z}-t) =
-e\vec{v}_{\perp}n_{0}(\vec{r}_{\perp})[1 +
a_{\mathrm{in}}\cos\omega(z/v_{z}-t)] \ ,
\end{displaymath}

\noindent where we calibrated the time in such a way that current
density has its maximum at time $t = 0$, at point $z = 0$. Even
through the measured quantities are real, it is generally more
convenient to use complex representation.  For this reason, starting
with real $\vec{j}_{\perp}$, one defines the complex transverse current
density:

\begin{equation}
j_{x} + \I j_{y} = -ec\theta_{\mathrm{w}}n_{0}(\vec{r}_{\perp})\exp(-\I
k_{\mathrm{w}}z) [1 + a_{\mathrm{in}}\cos\omega(z/v_{z}-t)] \ .
\label{eq:c}
\end{equation}

\noindent Transverse current have the angular
frequency $\omega$ and two waves  traveling in the same directions with
variations $\exp\I(\omega z/v_{z} - k_{\mathrm{w}}z - \omega t)$ and
$\exp-\I(\omega z/v_{z} + k_{\mathrm{w}}z- \omega t)$ will add to give
a total current proportional to $\exp(-\I
k_{\mathrm{w}}z)\cos\omega(z/v_{z}-t)$.  The factor $\exp\I(
\omega z/v_{z} - k_{\mathrm{w}}z - \omega t)$ indicates a fast wave,
while the factor $\exp-\I(\omega z/v_{z} + k_{\mathrm{w}}z - \omega t)$
indicates a slow wave. The use of the word "fast" ("slow") here implies
a wave with a phase velocity faster (slower) than the beam
velocity.

Now we should consider the electrodynamic problem. Using Maxwell's
equations, we can write the equation for the electric field:

\begin{displaymath}
c^{2}\vec{\nabla}\times(\vec{\nabla}\times\vec{E}) = -
\partial^{2}\vec{E}/\partial t^{2} - 4\pi
\partial\vec{j}/\partial t \ .
\end{displaymath}

\noindent Then make use of the identity

\begin{displaymath}
\vec{\nabla}\times(\vec{\nabla}\times\vec{E}) =
\vec{\nabla}(\vec{\nabla}\cdot\vec{E}) - \vec{\nabla}^{2}\vec{E} \ ,
\end{displaymath}

\noindent where $\vec{\nabla}\cdot\vec{E}$ can be found from the
Poisson equation. Finally, we come to the inhomogeneous wave equation
for $\vec{E}$:

\begin{equation}
c^{2}\vec{\nabla}^{2}\vec{E} -
\partial^{2}\vec{E}/\partial t^{2} =
4\pi c^{2}\vec{\nabla}\rho +
4\pi\partial\vec{j}/\partial t \ .
\label{eq:u1}
\end{equation}

\noindent This equation allows one to calculate the electric field
$\vec{E}(\vec{r},t)$ for given charge and current sources,
$\rho(\vec{r},t)$ and $\vec{j}(\vec{r},t)$. Thus, equation
(\ref{eq:u1}) is the complete and correct formula for radiation.
However we want to apply it to a still simpler circumstance in which
the second term (or, the current term) in the right-hand
side of (\ref{eq:u1}) provides the main contribution to the value of
the radiation field. Since in the paraxial
approximation the radiation field has only transverse components, we
are interested in the transverse component of (\ref{eq:u1}). Thus we
consider the wave equation

\begin{equation}
c^{2}\vec{\nabla}^{2}\vec{E}_{\perp} -
\partial^{2}\vec{E}_{\perp}/\partial t^{2} =
4\pi\partial\vec{j}_{\perp}/\partial t \ ,
\label{eq:u3}
\end{equation}

\noindent which relates the transverse component of the
electric field to the transverse component of current density.

We wish to examine the case when the phase
velocity of the current wave is close to the velocity of light.  This
requirement may be met under resonance condition

\begin{equation}
\omega/c = \omega/v_{z} - k_{\mathrm{w}} \ .
\label{eq:s}
\end{equation}

\noindent First we may point out that the statement of (\ref{eq:s}),
the condition for the relation between $\omega, k_{\mathrm{w}}$ and
$v_{z}$, is the condition for synchronism between the transverse
electromagnetic wave and the fast transverse current wave with the
propagating constant $\omega/v_{z} - k_{\mathrm{w}}$. With a current
wave traveling with the same phase speed as the electromagnetic wave,
we have the possibility of (space) resonance between electromagnetic
wave and electrons.  If this is the case cumulative interaction between
bunched electron beam and transverse electromagnetic wave takes place.
We are therefore justified in considering the contributions of all the
waves except the synchronous one to be negligible.

It would be nice to find an explanation of resonance condition
(\ref{eq:s}) which is elementary in the sense that we can see what is
happening physically.  Let us call the plane electromagnetic wave
velocity $c$, which is of course, related to the phase constant by $k =
\omega/c$.  Then the velocity of the beam relative to the
electromagnetic wave is merely $c-v_{z}$. With this relative velocity,
the frequency of the electromagnetic waves as seen by the electrons is
give by $\omega_{\mathrm{e}} = (c-v_{z})k$, where we have used the
obvious relation between the phase constant $k$ and the wavelength of
the electromagnetic wave, $k = 2\pi/\lambda$.  Relation for
$\omega_{\mathrm{e}}$ can be written as $\omega_{\mathrm{e}} = \omega -
kv_{z}$. Under the condition when (\ref{eq:s}) holds, we see by
comparison with that equation that this would be merely the undulator
frequency $\omega_{\mathrm{e}} = k_{\mathrm{w}}v_{z}$. Accordingly, the
condition of (\ref{eq:s}) that we have imposed is merely that the
electron velocity is such that as the electron run past the
electromagnetic waves, the apparent frequency they see is the undulator
frequency. This is the basic effect involved in this interaction. If
the apparent frequency is the undulator frequency, a cumulative effect
on the electromagnetic wave results, since the waves are being driven
at a frequency corresponding to natural rotation frequency of the
electrons in the magnetic field.  Another way of saying the same thing
is that in the frame of reference moving with the electrons, the
electromagnetic waves have a Doppler-shifted frequency equal to
undulator frequency.

Any state of transverse electromagnetic wave can always be written as a
linear combination of the two base states (polarizations).
By giving the amplitudes and phases of these base states we completely
describe the electromagnetic wave state.
It is usually best to start with the form which is physically clearest.
We choose the Cartesian base states and seek the solution for
$\vec{E}_{\perp}$ in the form

\begin{equation}
E_{x,y} = \tilde{E}_{x,y}(z,\vec{r}_{\perp})
\exp[\I\omega(z/c -t)]  + {\mathrm{C.C.}}
\label{eq:f}
\end{equation}

\noindent Here and in what follows, complex amplitudes related to the
field are written with a tilde. The description of the field given by
(\ref{eq:f}) is quite general. However, the usefulness of the concept
of carrier wave number is limited to the case where the amplitude is
slowly varying function of $z$.

To determine the form of
$\tilde{E}_{x,y} (z,\vec{r}_{\perp})$ we substitute (\ref{eq:c}) and
(\ref{eq:f}) into (\ref{eq:u3}). We have

\begin{eqnarray}
& \mbox{} &
\exp[\I\omega(z/c-t)]\left\{\vec{\nabla}^{2}_{\perp} +
\frac{2\I\omega}{c}\frac{\partial}{\partial z} +
\frac{\partial^{2}}{\partial z^{2}}\right\}{\tilde{E}_{x} \choose
\tilde{E}_{y}} + {\mathrm{C.C.}} \nonumber\\ & \mbox{} & = -
4\pi\frac{\omega}{c} {\cos(k_{\mathrm{w}}z) \choose
-\sin(k_{\mathrm{w}})} e\theta_{\mathrm{w}}a_{\mathrm{in}}
n_{0}(\vec{r}_{\perp})\sin\omega(z/v_{z}-t) \ .
\label{eq:u4a}
\end{eqnarray}

\noindent Here $\vec{\nabla}^{2}_{\perp}$ is the Laplace operator in
transverse coordinates.

Now we have an apparently simple pair of equations - and they
are still exact, of course. The question is, how to solve these
equations? First  we simplify the equations obtained by noting that for
a radiation field it is reasonable to assume that
$\tilde{E}_{x,y}(z,\vec{r}_{\perp})$ are slowly varying functions of
$z$ (because of the radiation directivity) so that
$\partial^{2}\tilde {E}_{x,y}/\partial z^{2}$ may be neglected.
The corresponding requirement for the complex amplitude is
$\mid
\partial\tilde{E}_{x,y}/\partial z\mid \ll k\mid\tilde
{E}_{x,y}\mid$. In other words, the radiation pulse must not change
significantly while traveling through a distance comparable with the
wavelength $\lambda = 2\pi/k$. This assumption is not a restriction.
Such is the case in all practical cases of interest. Differential
equations becomes

\begin{eqnarray}
& \mbox{} &
\exp[\I\omega(z/c-t)]\left\{\vec{\nabla}^{2}_{\perp} +
\frac{2\I\omega}{c}\frac{\partial}{\partial z}\right\}{\tilde{E}_{x}
\choose \tilde{E}_{y}} + {\mathrm{C.C.}}
\nonumber\\
& \mbox{} &
=
4\pi\frac{\omega}{c} {\cos(k_{\mathrm{w}}z) \choose
-\sin(k_{\mathrm{w}})} e\theta_{\mathrm{w}}a_{\mathrm{in}}
n_{0}(\vec{r}_{\perp})\sin\omega(t-z/v_{z}) \ .
\label{eq:u4}
\end{eqnarray}

\noindent The reader may well wonder why such a transformation is
useful; therefore we digress temporarily to address this question.

Such a transformation immediately modifies the hyperbolic wave
equations to the parabolic equations. We should stress that the
equations (\ref{eq:u4}) are preferable for an analytical solution,
since the mathematical techniques are always connected with more
conventional parabolic equations. Although equations (\ref{eq:u4})
cannot be solved in general, we will solve them for some special cases.
These equations can be simplified further by noting that
the complex amplitudes $\tilde{E}_{x,y}$ will not vary much with $z$,
especially in comparison with the exponential terms $\exp(-\I
k_{\mathrm{w}}z)$.  The slow wave of transverse current oscillates very
rapidly about an average value of zero and, therefore, does not
contribute very much to the rate of change of $\tilde{E}_{x,y}$.  So we
can make a reasonably good approximation by replacing these terms by
their average value, namely, zero. We will leave them out, and take as
our approximation:

\begin{equation}
\vec{\nabla}^{2}_{\perp}{\tilde{E}_{x}\choose \tilde{E}_{y}} +
\frac{2\I\omega}{c}\frac{\partial}{\partial
z}{\tilde{E}_{x} \choose \tilde{E}_{y}} = - {\I \choose 1}
2\pi\frac{\omega}{c}e\theta_{\mathrm{w}} a_{\mathrm{in}}
n_{0}(\vec{r}_{\perp})\exp(-\I Cz) \ .
\label{eq:u5}
\end{equation}

\noindent Even the remaining terms, with exponents proportional to $C =
\omega/v_{z} - \omega/c - k_{\mathrm{w}}$ will also vary rapidly
unless $C$ is near zero.  Only then will the right-hand side vary
slowly enough that any appreciable amount will accumulate when we
integrate the equations with respect to $z$. The required conditions
will be met if

\begin{displaymath}
C \ll k_{\mathrm{w}} \qquad 1 \ll k_{\mathrm{w}}z  \ .
\end{displaymath}

\noindent In other words, we use
the resonance approximation here and assume that complex amplitudes
$\tilde{E}_{x,y}$ are slowly varying in the longitudinal coordinate.
By "slowly varying" we mean that $\mid
\partial\tilde{E}_{x,y}/\partial z\mid \ll k_{\mathrm{w}} \mid\tilde
{E}_{x,y}\mid $. For this inequality to be satistied, the spatial
variation of $\tilde{E}_{x,y}$ within an undulator period
$\lambda_{\mathrm{w}} = 2\pi/k_{\mathrm{w}}$ has to be small.

Equations (\ref{eq:u5}) are simple enough and can be solved
in any number of ways. One convenient way is the following. Taking the
sum and the difference of the two we get

\begin{equation}
\left(\vec{\nabla}^{2}_{\perp} +
\frac{2\I\omega}{c}\frac{\partial}{\partial
z}\right)\left(\tilde{E}_{x} + \I \tilde{E}_{y}\right) =
2\pi\I\frac{\omega}{c} e\theta_{\mathrm{w}} a_{\mathrm{in}}
n_{0}(\vec{r}_{\perp})\exp(-\I Cz) \ ,
\label{eq:u6}
\end{equation}

\begin{equation}
\left(\vec{\nabla}^{2}_{\perp} +
\frac{2\I\omega}{c}\frac{\partial}{\partial
z}\right)\left(\tilde{E}_{x} - \I \tilde{E}_{y}\right) = 0 \ .
\label{eq:u7}
\end{equation}

\noindent These equations describe the general case of electromagnetic
wave radiation by the bunched electron beam in the
helical undulator. Equation (\ref{eq:u6}) and (\ref{eq:u7}) refer to
the right- and left-helicity components of the wave respectively.
The solutions for the right- and left-helicity waves are linearly
independent \footnote{ In the general case these components are not
independent, but are connected by the boundary conditions on the vacuum
chamber walls.  It is usually assumed that the vacuum chamber walls
are placed far enough from the electron beam, formally at infinity.
Such an approximation well describes FEL amplifiers operating in the
visible down to X-ray wavelength range. To describe FEL amplifier
operating in the far-infrared wavelength range one should take into
account the influence of the walls on the amplification process}.  It
follows from (\ref{eq:u6}) and (\ref{eq:u7}) that only those waves are
radiated that have the same helicity as the undulator field itself.

Of course we could predict such
a result. The electric field, $\vec{E}_{\perp}$, of the   wave radiated
in the helical undulator in resonance approximation is circularly
polarized and may be represent in the complex form:

\begin{equation}
E_{x} + \I E_{y} = \tilde{E}(z,\vec{r}_{\perp})
\exp[\I\omega(z/c -t)]  \ .
\label{eq:d}
\end{equation}

\noindent Finally, the equation for $\tilde{E}$ can be written in the
form

\begin{equation}
\left(\vec{\nabla}^{2}_{\perp} +
\frac{2\I\omega}{c}\frac{\partial}{\partial
z}\right)\tilde{E} =
2\pi\I\frac{\omega}{c} e\theta_{\mathrm{w}}a_{\mathrm{in}}
n_{0}(\vec{r}_{\perp})\exp(-\I Cz) \ .
\label{eq:u8}
\end{equation}

\noindent Equation (\ref{eq:u8}) is an inhomogeneous parabolic
equation. Its solution can be expressed in terms of a convolution
of the free-space Green's function (impulse response)

\begin{equation}
G(z-z^{\prime}, \vec{r}_{\perp}-\vec{r}^{\prime}_{\perp})
= \frac{1}{4\pi(z - z^{\prime})}
\exp\left[\frac{\I\omega\mid \vec{r}_{\perp} -
\vec{r}^{\prime}_{\perp}\mid^{2}}{2c(z- z^{\prime})}\right] \
\label{eq:ir}
\end{equation}

\noindent with the source
term. When the right-hand side of
(\ref{eq:u8}) is equal to zero, it transforms to the well-known
paraxial wave equation in optics.

The radiation process displays resonance behaviour and the amplitude of
electric field depends stronly on the value of the detuning
parameter $C$. With the approximation made in getting (\ref{eq:u8}) the
equation can be solved exactly, but the work is a little elaborate, so
we won't do that until later when we take up the problem of output
radiation characteristics.  Now we will find an exact solution for the
case of perfect resonance.  When the parameters
are tuned to perfect resonance, such that $C = 0$, the solution of the
equation (\ref{eq:u8}) has the form

\begin{equation}
\tilde{E}(z,\vec{r}_{\perp}) = \frac{\I
e\theta_{\mathrm{w}}\omega a_{\mathrm{in}}}{2c}
\int\limits^{z}_{0}\frac{\D z^{\prime}}{z -
z^{\prime}}
\int\D\vec{r}^{\prime}_{\perp}n_{0}(\vec{r}_{\perp}^{\prime})
\exp\left[\frac{\I\omega\mid \vec{r}_{\perp} -
\vec{r}^{\prime}_{\perp}\mid^{2}}{2c(z- z^{\prime})}\right] \ ,
\label{eq:r}
\end{equation}

\noindent where $(z,\vec{r}_{\perp})$ and
$(z^{\prime},\vec{r}_{\perp}^{\prime})$ are the coordinates of the
observation and the source point, respectively.

Let us consider an axisymmetric electron beam with gradient profile of
the current density. In this case we have
$-ev_{z}n_{0}(\vec{r}_{\perp}) = - j_{0}S(r)$, where $r$ is the radial
coordinate of the cylindrical system $(r,\phi,z)$ and $S(r)$ describes
the transverse profile of the electron beam. To be specific, we write
down all the expressions for the case of a Gaussian transverse
distribution:

\begin{displaymath}
-ev_{z}n_{0}(\vec{r}_{\perp}) = - j_{0}S(r) =
- \frac{I_{0}}{2\pi\sigma^{2}} \exp\left( -
\frac{r^{2}}{2\sigma^{2}}\right) \ ,
\end{displaymath}

\noindent where $I_{0}$ is the total beam current.
At this point we find it convenient to impose the following
restriction: we focus only on the radiation see by an observer lying
on the electron beam axis.
When $r = 0$ and the beam profile is Gaussian, we can write
(\ref{eq:r}) in the form

\begin{displaymath}
\tilde{E}_{0}(z) = \frac{\I
\theta_{\mathrm{w}}\omega a_{\mathrm{in}}I_{0}}{2\sigma^{2}c^{2}}
\int\limits^{z}_{0}\frac{\D z^{\prime}}{z -
z^{\prime}}
\int\limits^{\infty}_{0}r^{\prime}\exp\left[ -
\frac{(r^{\prime})^{2}}{2\sigma^{2}}\right]
\exp\left[\frac{\I\omega (r^{\prime})^{2}}{2c(z- z^{\prime})}\right]
\D r^{\prime} \ .
\end{displaymath}

\noindent Integrating first with respect to $r^{\prime}$, we have

\begin{equation}
\tilde{E}_{0}(z) = \frac{\I
\theta_{\mathrm{w}}\omega a_{\mathrm{in}}I_{0}}{2c^{2}}
\int\limits^{z}_{0}\frac{\D z^{\prime}}{z -
z^{\prime} +\I k\sigma^{2}} \ ,
\label{eq:r1}
\end{equation}

\noindent where $k = \omega/c$ is the radiation wavenumber.
The physical implications of this result are best understood by
considering some limiting cases. It is convenient to rewrite this
expression in a dimensionless form.  After appropriate normalization
it is a function of one dimensionless parameter only:

\begin{displaymath}
\hat{E}_{0} = f(\hat{z},\beta) = \I\int\limits^{\hat{z}}_{0}\frac
{\D\hat{z}^{\prime}}{\hat{z} - \hat{z}^{\prime} + \I\beta} \ ,
\end{displaymath}

\noindent where $\hat{z} = z/L_{\mathrm{w}}$ is the dimensionless
coordinate along the undulator, $L_{\mathrm{w}}$ is the total undulator
length, $\beta = k\sigma^{2}/L_{\mathrm{w}}$ is the diffraction
parameter and

\begin{displaymath}
\hat{E}_{0} = \tilde{E}/E_{\mathrm{n}} =
2c^{2}\tilde{E}/(\theta_{\mathrm{w}}\omega a_{\mathrm{in}}I_{0})
\end{displaymath}

\noindent is the normalized field amplitude. The
changes of scale performed during the normalization process, mean that
we are measuring distance along the undulator and electron beam size as
multiples of "natural" undulator radiation units. Let us study the
asymptotic behaviour of the field amplitude at large values of the
diffraction parameter $\beta$.  In this case $\beta \gg \hat{z} -
\hat{z}^{\prime}$  and we have asymptotically:

\begin{equation}
\hat{E}_{0} \to  \hat{z}/\beta \qquad {\mathrm{as}}\quad \beta \to
\infty \ .
\label{eq:a1}
\end{equation}

Now let us study the asymptote of a thin electron beam. In this case
$\beta \to 0$ and the function $f(\hat{z},\beta)$ can be evaluated in
the following way.  First we remark that the integral can be expressed
as $p + \I q$:

\begin{displaymath}
f(\hat{z},\beta) = \I\int\limits^{\hat{z}}_{0}\frac
{\D\hat{z}^{\prime}}{\hat{z} - \hat{z}^{\prime} + \I\beta}
= \int\limits^{\hat{z}/\beta}_{0}\frac{\D x}{1 + x^{2}} +
\I\int\limits^{\hat{z}/\beta}_{0}\frac{x\D x}{1 + x^{2}}
\end{displaymath}

\noindent The first integral should go from $0$ to $\hat{z}/\beta$, but
$0$ is so far from $\hat{z}/\beta$  that the curve is all
finished by that time , so we go instead to $\infty$  - it makes no
difference and it is much easier to do the integral. The integral is an
inverse tangent function. If we look it up in a book we see that it is
equal to $\pi/2$. The second integral can be expressed as logarithm
function. Thus we have

\begin{equation}
\hat{E}_{0} \to  \pi/2 + \I\ln(\hat{z}/\beta) \qquad {\mathrm{as}}\quad
\beta \to 0 \ .
\label{eq:a2}
\end{equation}

\noindent It is interesting to note
that in the case of a thin electron beam, the radiation field, acting
on the electrons is almost constant along the undulator axis, which is
a little strange, because the modulated electron beam radiates
electromagnetic waves along the undulator. But that is the way it comes
out - fortunately a rather simple formula \footnote{ For FEL experts
who happen to be reading this we should add that our logarithmic term
in (\ref{eq:a2}) and the logarithmic growth rate asymptote for
conventional FEL amplifier at small diffraction parameter (see
\cite{book}) are ultimately connected}.

Special attention is called to the fact that
in the thin electron beam case, at $\beta \to 0$, amplitude
$\tilde{E}(z)$ is a complex function. One immediately recognizes the
physical meaning of the complex $\tilde{E}_{0}(z)$. Note that electric
field (response) is given by fast wave of transverse current ("force")
times a certain factor.  This factor can either be written as $p + \I
q$, or as a certain magnitude $\rho$ times $\exp(\I\delta)$. If it is
written as a certain amplitude $\rho$ times $\exp(\I\delta)$, let
us see what it means.  This tells us that electric field is not
oscillating in phase with the fast wave of transverse current, which
has (at $C = 0$) the phase $ \psi = \omega z/c - \omega t$, but is
shifted by an extra amount $\delta(z)$.  Therefore $\delta(z)$
represent the phase shift of the response.

To get an intuitive picture on what happens with the radiation beam
inside the near zone according to equation (\ref{eq:r}), let us first
choose  thin beam asymptotic. This is an example in which diffraction
effects play an important role. Simple physical consideration can lead
directly to a crude approximation for the radiation beam cross-section.
There is a complete analogy between the radiation effects of
the bunched electron beam in the undulator and the radiation effects of
a sequence of periodically spaced oscillators.  The radiation of these
oscillators always interferes coherently at zero angle with respect to
the undulator axis. When all the oscillators are in phase there is
a strong intensity in the direction $\theta = 0$.
An interesting question is, where is the minimum? If we have a triangle
with a small altitude $r \simeq z\theta$ and a long base $z$, then the
diagonal $s$ is longer than the base.  The difference is $\Delta = s -
z \simeq r^{2}/2z \simeq z\theta^{2}/2$.  When $\Delta$ is equal to one
wavelength, we get a minimum.  Now why do we get a minimum when $\Delta
\simeq \lambda$?  Because the contributions of various oscillators are
then uniformly distributed in phase from $0$ to $2\pi$.  In the limit
of small size of the electron beam interference will be constructive
within an angle of about $\theta_{\mathrm{c}} \simeq
1/\sqrt{kz}$.

In the limit of large electron beam size,
the spread of the angle distribution can be found from a
two-dimensional Fourier transform of the field. The radiation field
across the electron beam may be presented as a superposition of plane
waves, all with the same wavenumber $k = \omega/c$. The value of
$k_{\perp}/k$ gives the sine of the angle between the $z$ axis and the
direction of propagation of the plane wave. In the paraxial
approximation $k_{\perp}/k = \sin\theta \simeq \theta$. We can expect
that the typical width of the angular spectrum should be of the order
$\theta_{\mathrm{c}} \simeq
(k\sigma)^{-1}$, a consequence of the reciprocal width relations of
the Fourier transform pair $\Delta k_{\perp}\sigma \simeq 1$.

\begin{figure}[tb]
\begin{center}
\epsfig{file=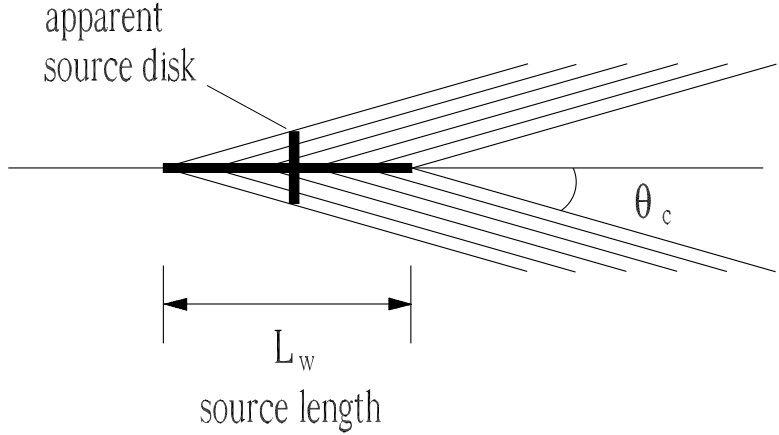,width=0.8\textwidth}
\end{center}
\caption{Apparent photon source size in the case of a thin electron
beam.  The photon source extends over undulator length $L_{\mathrm{w}}$
along the particle path} \label{fig:aps}
\end{figure}

The boundary between
these two asymptotes is about $k\sigma \simeq \sqrt{kL_{\mathrm{w}}}$
or (another way to write it) $\sigma ^{2}
\simeq \sigma_{\mathrm{dif}}^{2} = L_{\mathrm{w}}/k$.  A rough estimate
for the diffraction effects to be small is $\sigma \gg
L_{\mathrm{w}}/(k\sigma)$, which simply means that the diffraction
expansion of the radiation at undulator length must be much less than
the size of the beam.  Another way to write this condition is
$k\sigma^{2}/L_{\mathrm{w}} =  \beta \gg 1$. The parameter $\beta$ can
be referred to as the electron beam Fresnel number.

Our estimates
show (see Fig. \ref{fig:aps}) that in the case of a thin beam the
square of the radiation beam grows linearly with the undulator length

\begin{displaymath}
S_{\mathrm{rad}} \simeq (z\theta_{\mathrm{c}})^{2} \simeq z/k \ \
\ \ \qquad {\mathrm{for}} \quad \beta \ll 1 \ .
\end{displaymath}

\noindent In the case of a wide electron beam the most of the radiation
overlaps with the electron beam

\begin{displaymath}
S_{\mathrm{rad}} \simeq \sigma^{2} = {\mathrm{const.}} \qquad \ \ \
{\mathrm{for}} \quad 1 \ll \beta \ .
\end{displaymath}

Up to this point we only talked about electric field amplitude on the
electron beam axis. Ultimately we want to consider the distribution of
the electric field over the full electron beam cross-section. The full
analysis is a difficult analytical problem. The calculations can be
performed without great difficulty in two limiting cases, namely, the
cases of diffraction parameter very large and very small comared with
unity. To calculate equation (\ref{eq:r}) we note that the behaviour of
Green's function (\ref{eq:ir}) for $k\sigma^{2}/L_{\mathrm{w}} \gg 1$
approaches the behaviour of the delta function.  The source function
$n_{0}(\vec{r}^{\prime}_{\perp})$ does not vary very much across the
region $\mid\vec{r}_{\perp} - \vec{r}^{\prime}_{\perp}\mid^{2} \simeq
L_{\mathrm{w}}/k$ in the case of a wide electron beam:  therefore we
can replace it by a constant. In other words, we simply take
$n_{0}(\vec{r}^{\prime}_{\perp})$ outside the integral sign and call it
$n_{0}(\vec{r}_{\perp})$. In this case the integral over source
coordinates $\vec{r}^{\prime}_{\perp}$ in (\ref{eq:r}) is calculated
analytically

\begin{equation}
\tilde{E}(z,\vec{r}_{\perp}) = \pi e\theta_{\mathrm{w}}z a_{\mathrm{in}}
n_{0}(\vec{r}_{\perp}) =
\frac{\theta_{\mathrm{w}}za_{\mathrm{in}}I_{0}}{2c\sigma^{2}}
\exp\left(-\frac{r^{2}}{2\sigma^{2}}\right)
\quad {\mathrm{as}} \quad \beta \gg 1 \ .
\label{eq:w1}
\end{equation}

\noindent We emphasize the following features of this result.
At exact resonance $(C = 0)$ amplitude $\tilde{E}$ is a real function.
This tells us that electric field is oscillating in
phase with the fast wave of transverse current.

Let us turn next to the second case where $\beta \ll 1$. It should be
noted that dealing with klystron amplifier this situation is more
preferable than the previous one. In thin electron beam case there is
only a small variation in electric field amplitude over electron beam
cross section. This feature follows naturally from the fact that the
radiation expands out and electron beam becomes
thin with respect to the radiation beam. Using (\ref{eq:r}), we find
that the field amplitude inside the thin electron beam is
given by the formula

\begin{equation}
\tilde{E}(z,\vec{r}_{\perp}) =
\frac{\theta_{\mathrm{w}}\omega a_{\mathrm{in}}I_{0}}{2c^{2}}
\left[\frac{\pi}{2} +
\I\ln\left(\frac{cz}{\omega\sigma^{2}}\right)\right]
\qquad {\mathrm{as}} \quad \beta \ll 1 \ .
\label{eq:t1}
\end{equation}

\noindent The nonzero imaginary term is a manifestation
of the fact that electric field is not oscillating in phase with the
fast wave of transverse current.

In (\ref{eq:r}) we have the general solution of the problem of finding
the field produced by modulated current, and this is the first part of
the problem of the operation of the high gain klystron amplifier. The
next question which we must take up is the converse one: Given an
electric field of the type we have just considered, what is the
electronic motion in such a field?

\subsection{Self-interaction within a modulated electron beam moving in
an undulator}

Since, the wavelength of the radiation pulse is much less than either
the transverse electron beam size or the pulse
length it can be accurately represented by a wave traveling at
$\omega/k \simeq c$.  Further, since the electron also travels at
$v_{z} \simeq c$, the electron and the radiation remain essentially
overlapped during the transit through the undulator. Although the
radiation fields alone do not affect the electron trajectory, the
interaction of the transverse radiation electric field with the
transverse oscillations of the electron induced by the static undulator
magnetic field causes the electron to either gain or lose energy.

As simple example of the interaction
between electrons and electromagnetic wave in an undulator we shall
consider a circularly polarized plane electromagnetic wave propagating
parallel to the electron beam.
The field of the electromagnetic wave has only a transverse component,
so the energy exchange between the electron and the electromagnetic
wave is due to the transverse component of the electron velocity. The
rate of electron energy change is

\begin{displaymath}
\frac{\D{\cal E}}{\D t} = m_{\mathrm{e}}c^{2}\frac{\D\gamma}{\D t} =
- e\vec{v}_{\perp}\cdot\vec{E}_{\perp} \ ,
\end{displaymath}

\noindent where $\vec{E}_{\perp}$ is the vector of the electric field
of the wave:

\begin{displaymath}
\vec{E}_{\perp} = E\left\{\vec{e}_{x}\cos\left[\omega(z/c-t)\right] +
\vec{e}_{y}\sin\left[\omega(z/c-t)\right]\right\} \ .
\end{displaymath}

\noindent Remembering that $\D z = v_{z}\D t$ we find

\begin{eqnarray}
& \mbox{} &
\frac{\D{\cal E}}{\D z} =
- \frac{e}{v_{z}}\left(v_{x}E_{x} + v_{y}E_{y}\right)
\nonumber\\
& \mbox{} &
\simeq
-e\theta_{\mathrm{w}}E\left\{\cos(k_{\mathrm{w}}z)\cos\left[\omega(
z/c-t)\right] -
\sin(k_{\mathrm{w}}z)\sin\left[\omega(z/c-t)\right]\right\}
\nonumber\\
& \mbox{} &
=
- e\theta_{\mathrm{w}}E\cos\left[k_{\mathrm{w}}z + \omega(z/c-t)\right]
\nonumber\\
& \mbox{} &
=
- e\theta_{\mathrm{w}}E\cos\psi \ .
\label{eq:1a}
\end{eqnarray}

\noindent The phase $\psi$ has a simple physical interpretation and is
equal to the angle between the transverse velocity of the particle,
$\vec{v}_{\perp}$, and the vector of the electric field.
For effective energy exchange between the electron and the wave,
the scalar product $(e\vec{v}_{\perp}\cdot\vec{E}_{\perp})$ should be
kept nearly constant along the whole undulator length, i.e. a
synchronism should be provided. This resonance condition may be written
as

\begin{displaymath}
\D\psi = k_{\mathrm{w}}\D z + (\omega/c)\D z - \omega\D t = 0 \ .
\end{displaymath}

\noindent Remembering that $\D z = v_{z}\D t$, we have $k_{\mathrm{w}}
+ \omega/c - \omega/v_{z} = 0$. Or, since
$\lambda = 2\pi c/\omega$,

\begin{displaymath}
\lambda_{\mathrm{w}}/v_{z} = \lambda/(c-v_{z}) \ .
\end{displaymath}

\noindent Thus, we see that synchronism takes place when the wave
advances the electron beam by one wavelength at one undulator period.
This resonance condition is exactly equation (\ref{eq:s}), so we have
come full circle. Electromagnetic wave excitation and production of the
electron beam energy modulation are ultimately connected.

From this simple analysis it is apparent that the electron's relative
position within a radiation wavelength will determine whether it
consistently gains or loses energy as it travels through the undulator
magnet. A convenient variable for the description of the interaction
between electrons and electromagnetic wave in an undulator is the phase
$\psi$.  The relevant value of the phase $\psi$ is that at the location
of the particle.  Hence, the total derivative of $\psi$ is given by

\begin{displaymath}
\frac{\D\psi}{\D z} =  C = \frac{\partial\psi}{\partial z}
+ \frac{\partial\psi}{\partial t}\frac{\D t}{\D z} = k_{\mathrm{w}} +
\frac{\omega}{c} - \frac{\omega}{v_{z}} \ .
\end{displaymath}

\noindent Thus, the phase of the particle changes when the resonance
condition is not satisfied exactly.

Since discrete electrons are considered, their entrance into
the interaction region is conveniently described in terms of their
entrance phase positions relative to one cycle of the modulating
electromagnetic wave at the input, i.e. $z = 0$.  An entrance phase
variable is defined by $\psi_{0} = - \omega t_{0}$, where $t_{0}$ is
the entrance time.  Let the initial electron energy  be ${\cal
E}_{0}$. Thus, at a given displacement plane the electron will have a
kinetic energy deviation

\begin{displaymath}
{\cal E}(\psi_{0},z) - {\cal E}_{0}  = -
e\theta_{\mathrm{w}}\int\limits^{z}_{0}\D z
E\cos\left(\psi_{0} + Cz \right) \ ,
\end{displaymath}

\noindent if $\psi_{0}$ is the arrival phase of this particular
electron at the undulator entrance.

We should remark that our analysis of the interaction between electrons
and radiation in an undulator gives a result that is
somewhat simpler than we would actually find in nature.
All of our calculations
have been made for the plane electromagnetic
wave with phase velocity equal to the velocity of light. The actual
radiation wave in an undulator will experience an amplitude change and
a phase shift due to the radiation process.  The electric field of the
wave radiated in the helical undulator can be represented by equation
(\ref{eq:d}).  We can also write this expression as

\begin{displaymath}
\tilde{E}(z,\vec{r}_{\perp})\exp\I\omega(z/c-t) = \mid
\tilde{E}(z,\vec{r}_{\perp})\mid\exp\I[\omega(z/c-t) +
\psi_{\mathrm{e}}(z,\vec{r}_{\perp})] \ ,
\end{displaymath}

\noindent which says that radiation wave is not ordinary transverse
plane wave. Our final
result for the kinetic energy deviation is therefore

\begin{equation}
{\cal E} - {\cal E}_{0}  = -
e\theta_{\mathrm{w}}\int\limits^{z}_{0}\D z
\mid\tilde{E}(z,\vec{r}_{\perp})\mid\cos\left[\psi_{0} + Cz +
\psi_{\mathrm{e}}(z,\vec{r}_{\perp})\right] \ ,
\label{eq:m1}
\end{equation}

This expression deserves some remarks.
In dealing with start-up from the electron beam density modulation the
appropriate "frame of reference" is moving with the fast wave
rather than at the velocity of light.
The angle between the transverse velocity of the particle
and the vector of the electric field is given by

\begin{displaymath}
\psi = \psi_{0} + Cz +
\psi_{\mathrm{e}}(z,\vec{r}_{\perp}) \ ,
\end{displaymath}

\noindent where $Cz + \psi_{\mathrm{e}}(z,\vec{r}_{\perp})$ represent
the phase shift of the radiation wave ("response") relative to the fast
wave ("force").  What do we use for the entrance
phase in our formula (\ref{eq:m1})? We use the entrance phase position
relative to one cycle of the  fast wave at the input. For simplicity
of analysis,  we consider the cases in which there are no variations in
phase of the density modulation in the transverse plane at the
undulator entrance. In those cases an
entrance phase variable is defined by $\psi_{0} = - \omega t_{0}$,
where $t_{0}$ is the  entrance time.

We have what we need to know - the electron beam energy
modulation at the undulator exit (\ref{eq:m1}). We are ready to find
energy modulation, because we have already worked out what field is
produced by a bunched electron beam inside the near zone (\ref{eq:r}).
These equations are not well suited for quick
calculation of the energy modulation at a particular diffraction
parameter.  We may, however, express (\ref{eq:r}) in much simpler form
for very small and very large diffraction parameters, making use of
limiting expressions (\ref{eq:w1}) and (\ref{eq:t1}). Let us see what
happens if the diffraction parameter $\beta$ is large.  At exact
resonance $(C = 0)$ using (\ref{eq:w1}) the energy modulation amplitude
achieved at a given displacement plane can be written as:

\begin{equation}
\frac{\delta{\cal E}}{{\cal E}} =
- \frac{a_{\mathrm{in}}\theta^{2}_{\mathrm{w}}}
{4}\frac{z^{2}}{\sigma^{2}}\frac{I_{0}}
{\gamma I_{\mathrm{A}}}
\exp\left( - \frac{r^{2}}{2\sigma^{2}}
\right)\cos\psi_{0} \ ,
\label{eq:me1}
\end{equation}

\noindent where $I_{\mathrm{A}} = m_{\mathrm{e}}c^{3}/e \simeq 17
{\mathrm{kA}}$ is the Alfven current.

Now let us analyze this expression and study some of its consequences.
To develop a useful mathematical formalism for the description of
interaction between electrons and fields in an undulator we must choose
the appropriate coordinate system to maximize physical clarity. We
chose at the beginning of this section to use the stationary coordinate
system $z$.  For example, the beam density at point $z$, at time $t$ is

\begin{displaymath}
n(z/v_{z}-t) = n_{0} +
a_{\mathrm{in}}n_{0}\cos\omega(z/v_{z}-t) .
\end{displaymath}

\noindent A convenient way of description the particle beam dynamics is
to use a coordinate system $z^{\prime}$ that moves at a velocity
$v_{z}$ with respect to the stationary coordinate system $z$.  When a
particular electron enters the undulator, its initial position is
$z^{\prime}_{0}$ in the moving coordinate system.  The electron would
remain stationary at $z^{\prime}_{0}$ in the moving coordinate system;
in the stationary coordinate system, at a time $t$, it would have moved
a distance from the entrance $z = v_{z}(t-t_{0})$, where $t_{0}$ is the
entrance time.  Let us write an expression for beam density modulation
in terms of the moving coordinate system introducing the wave number
$k^{\prime}$ defined by $k^{\prime} = \omega/v_{z} =
2\pi/\lambda^{\prime}$,  where $\lambda^{\prime}$ is a wavelength of
the modulation frequency, i.e.  the distance through which a particle
travels in a cycle of modulation frequency. We may thus write
$n(z^{\prime}_{0}) = n_{0} + a_{\mathrm{in}}n_{0}
\cos\left(k^{\prime}z^{\prime}_{0}\right)$. Note that
$k^{\prime}z^{\prime}_{0} = - \omega t_{0} = \psi_{0}$ and  this
expression may be rewritten in the terms of arrival phase

\begin{displaymath}
n(\psi_{0}) = n_{0} + a_{\mathrm{in}}n_{0}
\cos\psi_{0} \ ,
\end{displaymath}

\noindent where we calibrated time in such a way that when a particular
electron enters the undulator at time $t = 0$, its initial position is
$z^{\prime} = 0$ in the moving coordinate system. Just as we expected,
in the case of wide electron beam, the beam energy modulation is
oscillating in phase with beam density modulation.

In the case of thin electron beam, things are quite different.
For a small diffraction parameter $\beta \ll 1$ we may apply an
asymptotic approximation for the field amplitude (\ref{eq:t1}) and get

\begin{equation}
\frac{\delta{\cal E}}{{\cal E}} =
- \frac{a_{\mathrm{in}}\theta^{2}_{\mathrm{w}}\omega z}
{2c}\frac{I_{0}}{\gamma I_{\mathrm{A}}}
\left\{\frac{\pi}{2}\cos\psi_{0} -
\left[\ln\left(\frac{cz}{\omega\sigma^{2}}\right)
- 1\right]\sin\psi_{0}  \right\} \ .
\label{eq:me2}
\end{equation}

\noindent Thus, we see that, in the case of thin electron
beam, the beam energy modulation is not oscillating in phase with beam
density modulation. Equations (\ref{eq:me1})  and (\ref{eq:me2}) give
us an idea of what we should expect. Generally we can try to
calculate the energy modulation precisely by using the (\ref{eq:r}) and
(\ref{eq:m1}). That is the end of our calculations of the electron
beam energy modulation, but there is one physically interesting thing
to check, and that is the conservation of energy.

\subsection{Power balance}

As any oscillating charge radiates energy, so must a
modulated electron beam moving along an undulator radiate energy. If
the system radiates energy, then in order to preserve
conservation of energy we must find that electron beam energy is being
lost.  An interesting question is, what forces are electrons working
against? That is interesting question which can be
completely and satisfactorily answered for modulated electron beam in
an undulator.  What happens in this:  in bunched beam, the fields
produced by the moving charges in one part of the undulator react on
the moving charges in another part of the undulator. We can calculate
these forces and find out how much work they do, and so find the right
rule for the radiation force.  We can make this calculation because at
short distance we know the electric field.  Above we calculated
the radiation field inside the near zone (\ref{eq:r}).

In our system energy is transported both in the form of electromagnetic
waves and in kinetic energy. The well-known Poynting vector represents
the electromagnetic power flow; the kinetic-power flow is merely
the number of electrons crossing per unit area per unit time
multiplied by the kinetic energy per electron. Consider first the
electromagnetic power.  In the paraxial approximation the diffraction
angles are small, the vectors of the electric and magnetic field are
equal in absolute value and are perpendicular to each other.  Thus, the
expression for the radiation power, $W$, can be written in the form:

\begin{displaymath}
W = \frac{c}{4\pi}\int\overline{\mid
\vec{E}_{\perp}\mid^{2}}\d\vec{r}_{\perp} \ ,
\end{displaymath}

\noindent where $\overline{(\cdots)}$ denotes averaging over a
cycle of oscillation of the carrier wave.  If we consider a
system with fields and bunched electron beam in an undulator, the
energy stored in any volume fluctuates sinusoidally with time. But
on the average there is no increase or decrease in the energy stored in
any portion of the volume, so that the below conservation theorem holds
if averaged over time, although the integrals might not be zero
instantaneously because of the time variations in the energy storage.

Since the radiation field inside the near zone has the form of equation
(\ref{eq:r}), than $W$ is given by:

\begin{eqnarray}
& \mbox{} &
W = \frac{c}{4\pi}\int\overline{\mid
\vec{E}_{\perp}\mid^{2}}\D\vec{r}_{\perp} =
\frac{c}{4\pi}\int\mid\tilde{E}(z,\vec{r}_{\perp})\mid^{2}
\D\vec{r}_{\perp}
\nonumber\\
& \mbox{} &
=
\frac{e^{2}\omega^{2}\theta^{2}_{\mathrm{w}}a^{2}_{\mathrm{in}}}{16\pi
c}\int\D\vec{r}_{\perp}
\left\{\int\limits^{z}_{0}\frac{\D z^{\prime}}{z-z^{\prime}}
\int\D\vec{r}_{\perp}^{\prime}n_{0}(\vec{r}^{\prime}_{\perp})
\exp\left[\frac{\I\omega\mid\vec{r}_{\perp}-\vec{r}^{\prime}_{\perp}
\mid^{2}}{2c(z-z^{\prime})}\right]\right\}
\nonumber\\
& \mbox{} &
\times
\left\{\int\limits^{z}_{0}\frac{\D z^{\prime\prime}}{z-z^{\prime\prime}}
\int\D\vec{r}_{\perp}^{\prime\prime}n_{0}(\vec{r}^{\prime\prime}_{\perp})
\exp\left[-
\frac{\I\omega\mid\vec{r}_{\perp}-\vec{r}^{\prime\prime}_{\perp}
\mid^{2}}{2c(z-z^{\prime\prime})}\right]\right\} \ .
\label{eq:b1}
\end{eqnarray}

\noindent The product of integrals over $z^{\prime}$ and
$z^{\prime\prime}$ can be represent as

\begin{displaymath}
\int\limits^{z}_{0}\Phi(z^{\prime})\D z^{\prime}\int
\limits^{z}_{0}\Phi^{*}(z^{\prime\prime})\D z^{\prime\prime}
=
\int\limits^{z}_{0}\Phi(z^{\prime})\D z^{\prime}\int
\limits^{z^{\prime}}_{0}\Phi^{*}(z^{\prime\prime})\D z^{\prime\prime}
+ {\mathrm{C.C.}}
\end{displaymath}

\noindent The integral over transverse coordinate $\vec{r}_{\perp}$ is
equal to

\begin{eqnarray}
& \mbox{} &
\int\D\vec{r}_{\perp}
\exp\left[
\frac{\I\omega\mid\vec{r}_{\perp}-\vec{r}^{\prime}_{\perp}
\mid^{2}}{2c(z-z^{\prime})} -
\frac{\I\omega\mid\vec{r}_{\perp}-\vec{r}^{\prime\prime}_{\perp}
\mid^{2}}{2c(z-z^{\prime\prime})}\right]
\nonumber\\
& \mbox{} &
=
\int\limits^{\infty}_{-\infty}\D x\int\limits^{\infty}_{-\infty}\D y
\exp\left\{\frac{\I\omega}{2c}\frac{[(x-x^{\prime})^{2} +
(y-y^{\prime})^{2}]}{z-z^{\prime}}
\right.
\nonumber\\
& \mbox{} &
\left.
-
\frac{\I\omega}{2c}\frac{[(x-x^{\prime\prime})^{2} +
(y-y^{\prime\prime})^{2}]}{z-z^{\prime\prime}}\right\}
\nonumber\\
& \mbox{} &
=
\frac{2\pi\I c}{\omega}\frac{(z-z^{\prime})(z-z^{\prime\prime})}
{z^{\prime}-z^{\prime\prime}}\exp\left[-\frac{\I\omega\mid\vec{r}^{\prime}
_{\perp} - \vec{r}^{\prime\prime}_{\perp}\mid^{2}}{2c(z^{\prime}
- z^{\prime\prime})}\right] \ .
\label{eq:b2}
\end{eqnarray}

\noindent As a result, expression (\ref{eq:b1}) can be written in the
form:

\begin{eqnarray}
& \mbox{} &
W = \frac{\I\omega e^{2}\theta^{2}_{\mathrm{w}}a^{2}_{\mathrm{in}}}
{8}\int\limits^{z}_{0}\D z^{\prime}\int\limits^{z^{\prime}}_{0}
\frac{\D z^{\prime\prime}}{z^{\prime} - z^{\prime\prime}}\int\D
\vec{r}^{\prime}_{\perp}\int\D\vec{r}^{\prime\prime}_{\perp}
n_{0}(\vec{r}^{\prime}_{\perp})n_{0}(\vec{r}^{\prime\prime}_{\perp})
\nonumber\\
& \mbox{} &
\times
\exp\left[- \frac{\I\omega\mid\vec{r}^{\prime}_{\perp} -
\vec{r}^{\prime\prime}_{\perp}\mid^{2}}{2c(z^{\prime}-z^{\prime\prime})}
\right] + {\mathrm{C.C.}}
\label{eq:b3}
\end{eqnarray}

\noindent Latter expression can be written in terms of the complex
amplitude of the radiated field $\tilde{E}(z,\vec{r}_{\perp})$. Using
(\ref{eq:r}) and (\ref{eq:b3}) we obtain the expression for $W$:

\begin{equation}
W = \frac{\I
e\theta_{\mathrm{w}}ca_{\mathrm{in}}}{4}\int\limits^{z}_{0}\D
z^{\prime}\int\D\vec{r}^{\prime}_{\perp}
n_{0}(\vec{r}^{\prime}_{\perp})
\tilde{E}(z^{\prime},\vec{r}^{\prime}_{\perp})
+ {\mathrm{C.C.}}
\label{eq:b4}
\end{equation}

The radiation power, $W$, must be equal to the difference of the
electron beam power at the exit and the entrance of the undulator. The
rate of the energy change for a single electron is given by

\begin{displaymath}
\D{\cal E}/\D z = - e\theta_{\mathrm{w}}
\mid\tilde{E}(z,\vec{r}_{\perp})\mid\cos[\psi_{0} +
\psi_{\mathrm{e}}(z,\vec{r}_{\perp})] \ .
\end{displaymath}

\noindent To obtain the mean power loss by the electron beam, we should
multiply this value by the particle flux density
$v_{z}a_{\mathrm{in}}n_{0}\cos\psi_{0} \simeq
ca_{\mathrm{in}}n_{0}\cos\psi_{0}$ average over time,
and integrate over the beam cross-section and the undulator length.
Finally, we obtain the following result:

\begin{eqnarray}
& \mbox{} &
\Delta W_{\mathrm{e}} = \frac{1}{2\pi}\int
\limits^{2\pi}_{0}\D\psi_{0}
\int\limits^{z}_{0}\D z^{\prime}\int\D\vec{r}^{\prime}_{\perp}
\left\{ca_{\mathrm{in}}n_{0}(\vec{r}^{\prime}_{\perp})
\cos\psi_{0}\right\}
\nonumber\\
& \mbox{} &
\times
\left\{-e\theta_{\mathrm{w}}\mid
\tilde{E}(z^{\prime},\vec{r}^{\prime}_{\perp}) \mid
\cos[\psi_{0} +
\psi_{\mathrm{e}}(z^{\prime},\vec{r}^{\prime}_{\perp})]\right\}
\nonumber\\ & \mbox{} & = - \frac{\I
e\theta_{\mathrm{w}}ca_{\mathrm{in}}}{4}\int\limits^{z}_{0}\D
z^{\prime}\int\D\vec{r}^{\prime}_{\perp}
n_{0}(\vec{r}^{\prime}_{\perp})
\tilde{E}(z^{\prime},\vec{r}^{\prime}_{\perp})
+ {\mathrm{C.C.}}
\label{eq:b5}
\end{eqnarray}

\noindent Comparing this expression with (\ref{eq:b4}), we see that the
radiation power and the change in the electron beam power have equal
absolute values and are opposite in sign, i.e. $\Delta W_{\mathrm{e}} +
W = 0$.  So, power balance takes place.

\section{The region of applicability}

Our theory of klystron amplifier is based on the assumption that beam
density modulation does not appreciably change as the beam propagates
through the undulator. When the resonance condition takes place, the
electrons with different arrival phases acquire different values of the
energy increments (positive or negative), which results in the
modulation of the longitudinal velocity of the electrons $v_{z}$ with
the radiation frequency $\omega$. Since this velocity modulation is
transformed into density modulation of the electron beam during the
undulator pass, an additional radiation field exists because of
variation in amplitude density modulation. Instead, we assume that the
amplitude of the electron beam density modulation has the same value at
all points in the undulator.  This approximation means that only the
contributions to the radiation field arising from the initial density
modulation are taken into account, and not those arising from the
induced bunching.

\subsection{Induced bunching}

It is interesting to estimate
the amount of bunching produced during the undulator pass. The velocity
$v_{z}$ of the electron at given displacement plane $z$ is given by

\begin{displaymath}
v_{z}({\cal E}) = v_{0} + \delta
v_{z}(z,\vec{r}_{\perp})\cos[\psi_{0} + \delta(z,\vec{r}_{\perp})]
\end{displaymath}

\noindent where $v_{0} = v_{z}({\cal E}_{0})$ is the injected
velocity, and $\delta v_{z}$ is a small amplitude of high frequency
oscillation.  From (\ref{eq:me2}) we can find $\delta v_{z}$
remembering that $v_{z} \simeq c$ and $(\D v_{z}/\D{\cal E})\mid_{{\cal
E}={\cal E}_{0}} \simeq c/(\gamma^{2}_{z}{\cal E}_{0})$. We have

\begin{equation}
\delta v_{z}(z) \simeq
a_{\mathrm{in}}\theta^{2}_{\mathrm{w}}\omega zI_{0}
\left(\gamma^{2}_{z}\gamma I_{\mathrm{A}}\right)^{-1} \ .
\label{eq:dv}
\end{equation}

\noindent We now examine the behaviour of these electrons as they
travel a certain distance in an undulator. Since successive electrons
are not traveling with the same velocity,
bunching takes place. To calculate a numerical value for the current,
we must compute the actual arrival time of the various electrons at
the undulator exit. The appropriate expression is formulated by means
of the law of conservation of charge. Quantitatively, we can say that
during an infinitesimal interval of time $\D t_{1}$, a total charge $\D
q = I_{1}\D t_{1}$ passes through plane $z_{1}$. We follow the
same particles (i.e. the same charge) through the plane
$z_{2}$ at some later time $t_{2}$.  These particles will pass through
the $z_{2}$ plane during the an infinitesimal time interval $\D t_{2}$
which may be shorter or longer than $\D t_{1}$. In either case, the
instantaneous current $I_{2}(t_{2})$ due to the charge passing through
$z_{2}$ plane in the interval $\D t_{2}$ is obtained from $\D q =
I_{2}\D t_{2}$. Obviously, the two charges are the same since we are
following the same set of particles.  Hence due to this charge, there
is a current through $z_{2}$ plane $I_{2} = I_{1}\D t_{1}/\D t_{2}$. If
an electron has velocity $v_{z}$ when it leaves point $z_{1}$ and
maintains this velocity, then its transit time to point $z_{2}$ is
given by the expression

\begin{equation}
t_{2} = t_{1} + \frac{z_{2}-z_{1}}{v_{0} + \delta v_{z}\cos\omega t_{1}
} \simeq t_{1} + \frac{z_{2} - z_{1}}{v_{0}}\left[ 1 - \frac{\delta
v_{z}}{v_{0}}\cos\omega t_{1}\right] \ .
\label{eq:ret}
\end{equation}

\noindent This
equation provides us with the relation between the arrival time $t_{2}$
at point $z_{2}$ for the particular group of charges which left point
$z_{1}$ at the departure time $t_{1}$.  It is possible to calculate
current at point $z_{2}$ for small values of $\delta v_{z}/v_{0}$  by
differentiating the last equation and substituting in equation $I_{2} =
I_{1}\D t_{1}/\D t_{2}$. This procedure leads to

\begin{displaymath}
I_{2} \simeq \frac{I_{1}}{1 + \alpha\sin\omega t_{1}} \simeq
I_{1}\left(1 - \alpha\sin\omega t_{1}\right) \ ,
\end{displaymath}

\noindent where $\alpha = \omega(z_{2}-z_{1})\delta v_{z}/v^{2}_{0}$.
This equation is obviously still not in a useful form, since
it also only gives the current at point $z_{2}$ in terms of departure
times of the charges from point $z_{1}$. However, by using
(\ref{eq:ret}) again (to eliminate $t_{1}$) and using the fact that
$\alpha \ll 1$, one can write approximately

\begin{displaymath}
I_{2} \simeq I_{1}\left\{ 1 - \alpha\sin\left[\omega t_{2} -
k\left(z_{2}-z_{1}\right)\right]\right\} \ ,
\end{displaymath}

\noindent where the terms that have been neglected in $t_{1}$ would
have contributed only terms of the order of $\alpha^{2}$ to the current
$I_{2}$. Thus the induced bunching that we want at the point $z_{2}$
due to the velocity modulation at the point $z_{1}$ is equal to $\alpha
= \omega(z_{2}-z_{1})\delta v_{z}/v^{2}_{0}$. Now we find the total
induced bunching at the point $z_{2}$ by finding the induced bunching
there from each point $z_{1} < z_{2}$, and then adding the
contributions from all the points.  To calculate this sum we need to
use (\ref{eq:dv}). We have, then,

\begin{equation}
\delta a =
\frac{\omega}{v^{2}_{0}}\int\limits^{z_{2}}_{0}\left(z_{2}-z_{1}\right)
\frac{\D v_{z}}{\D z_{1}}\D z_{1} \simeq
a_{\mathrm{in}}\omega^{2}\theta^{2}_{\mathrm{w}}z^{2}_{2}I_{0}
\left(c^{2}\gamma^{2}_{z}\gamma I_{\mathrm{A}}\right)^{-1} \ .
\label{eq:da}
\end{equation}

\noindent Thus, the requirement for the induced bunching to be small
can be written as $\delta a \ll a_{\mathrm{in}}$.  The result of a more
careful analysis shows that the last condition can be written as

\begin{equation}
\omega^{2}\theta^{2}_{\mathrm{w}}L^{2}_{\mathrm{w}}I_{0}
\left(6c^{2}\gamma^{2}_{z}\gamma I_{\mathrm{A}}\right)^{-1} \ll 1 \ .
\label{eq:c1}
\end{equation}

\subsection{Energy spread}

During the passage through an undulator the electron density modulation
can be suppressed by the longitudinal velocity spread in the electron
beam. For effective operation of klystron amplifier the value
of suppression factor should be close to unity.
If there is an (local) electron energy spread into the electron
bunch, $\Delta {\cal E}$, there will be a corresponding longitudinal
velocity spread given by $c\Delta{\cal E}/(\gamma^{2}_{z}{\cal
E}_{0})$.  To get a rough idea of the spread of electron density
modulation, the position of the particles within the electron beam at
the undulator exit has a spread which is equal to

\begin{displaymath}
\Delta z^{\prime} \simeq L_{\mathrm{w}}\Delta
v_{z}/v_{0} \simeq L_{\mathrm{w}}\Delta{\cal
E}\left(\gamma^{2}_{z}{\cal E}_{0}\right)^{-1}  \ ,
\end{displaymath}

\noindent We know that uncertainty in the phase of the
particles is about $\Delta\psi_{0} \simeq  2\pi\Delta
z^{\prime}/\lambda^{\prime}$. Therefore, a rough estimate for the
microbunching spread to be small is

\begin{displaymath}
\omega L_{\mathrm{w}}\Delta{\cal E}\left(c\gamma^{2}_{z}
{\cal E}_{0}\right)^{-1} \ll 1 \ .
\end{displaymath}

It is clear that we have made several rough approximations. Besides
various factors of two we have left out, we have used $\Delta{\cal E}$,
where we should have used variance $\langle\left(\Delta{\cal
E}\right)^{2}\rangle$. All of these refinements can be made; the
result of a more careful analysis shows that for the Gaussian
distribution function the suppression factor is equal to $\exp\left[-
\omega^{2}L_{\mathrm{w}}^{2} \langle\left(\Delta{\cal
E}\right)^{2}\rangle/\left(2c^{2}\gamma^{4}_{z} {\cal
E}_{0}^{2}\right)\right]$. So a better answer is

\begin{displaymath}
\frac{\omega^{2}L^{2}_{\mathrm{w}}}{2c^{2}\gamma^{4}_{z}}
\frac{\langle\left(\Delta{\cal E}\right)^{2}\rangle}{{\cal E}^{2}_{0}}
\ll 1 \ .
\end{displaymath}

\noindent Remembering that $\lambda_{\mathrm{w}} =
2\gamma^{2}_{z}\lambda$ and $v_{0} \simeq c$,  we have

\begin{equation}
(2\pi N_{\mathrm{w}})^{2}\langle\left(\Delta{\cal
E}\right)^{2}\rangle/{\cal E}^{2}_{0} \ll 1 \ ,
\label{eq:c2}
\end{equation}

\noindent where $N_{\mathrm{w}}$ is the number of undulator periods.
This condition is not a restriction.  Let us present a specific
numerical example. For unbunched electron beam the local energy spread
is about $\sqrt{\langle\left(\Delta{\cal E}\right)^{2}\rangle} \simeq \
5 \ {\mathrm{keV}}$. If the nominal energy is ${\cal E}_{0} = 500 \
{\mathrm{MeV}}$, the last condition can be written as $N_{\mathrm{w}}
\ll 10^{4}$. Such is the case in all practical cases of interest.

\subsection{Transverse emittance}

For a klystron amplifier it is of great interest to maximize the gain
per cascade pass which is proportional to the amplitude of electric
field inside the electron beam. It is usually desired to optimize the
amplifier to achieve the highest possible gain coefficient for a given
total current $I_{0}$.  Reducing the particle beam cross-section  by
diminishing the betatron function reduces also the size of the
radiation beam and increases, according to (\ref{eq:w1}), the amplitude
of the electric field.  This process of reducing the beam cross section
is, however, effective only up to some point.  Further reduction of the
particle beam size would have practically no effect on the radiation
beam size because of diffraction effects (\ref{eq:t1}).  It is
therefore prudent not to push the particle beam size to values much
less than the difraction limited radiation beam size.  From the
preceding discussion we may want to optimize the beam geometry as
follows. For maximum amplification, the transverse size of the electron
beam has to be matched to the diffraction limited radiation beam size

\begin{displaymath}
\sigma < \sqrt{\lambda L_{\mathrm{w}}/(2\pi)} \ .
\end{displaymath}

\noindent The wavelength and undulator length dictates the choice of
the optimal transverse size of the electron beam.

One might think that all we have to do is
to get electric field amplitude to a maximum - we can always reduce
electron beam cross section and we can always increase gain.  So it is
not impossible to build an electron focusing system that has a small
betatron function. In fact, one of the main problems in klystron
amplifier operation is preventing the spread of microbunching due to
angle spread into the electron beam. Some electrons traverse the
undulator not along or parallel to the $z-$axis, but at a small angle
$\theta$.  If there is an rms angular divergence $\sigma^{\prime}$
within the electron bunch there will be a corresponding longitudinal
velocity spread given by $\Delta v_{z}/v_{z} \simeq
(\sigma^{\prime})^{2}/2$.  We assume that the amplitude of beam
modulation  does not appreciably change as the beam propagates through
the undulator.  To avoid an decrease in the amplitude of the beam
modulation, the longitudinal velocity spread  must be small enough
$L_{\mathrm{w}}\Delta v_{z}/v_{z} \ll \lambda/(2\pi)$.  This gives the
condition

\begin{displaymath}
(\sigma^{\prime})^{2} \ll \lambda/(2\pi L_{\mathrm{w}}) \ .
\end{displaymath}

\noindent The last two conditions can be written as

\begin{displaymath}
\epsilon/\pi = \sigma\sigma^{\prime} \ll \lambda/(2\pi) \ ,
\end{displaymath}

\noindent where $\epsilon$ is the transverse electron beam emittance.
The last condition can be seen as "phase matching" of the electrons and
radiation. To produce maximal gain the particle beam emittance must be
less than the diffraction limited radiation emittance.  Obviously, this
condition is easier to achieve for long wavelengths.  For optimal gain
for a 30 nm radiation, for example, the electron beam emittance
must be smaller than about $10^{-6}\pi \ {\mathrm{rad.-cm.}}$
This condition may be easily satisfied in practice.

\section{Bunching by energy modulation}

\begin{figure}[tb]
\begin{center}
\epsfig{file=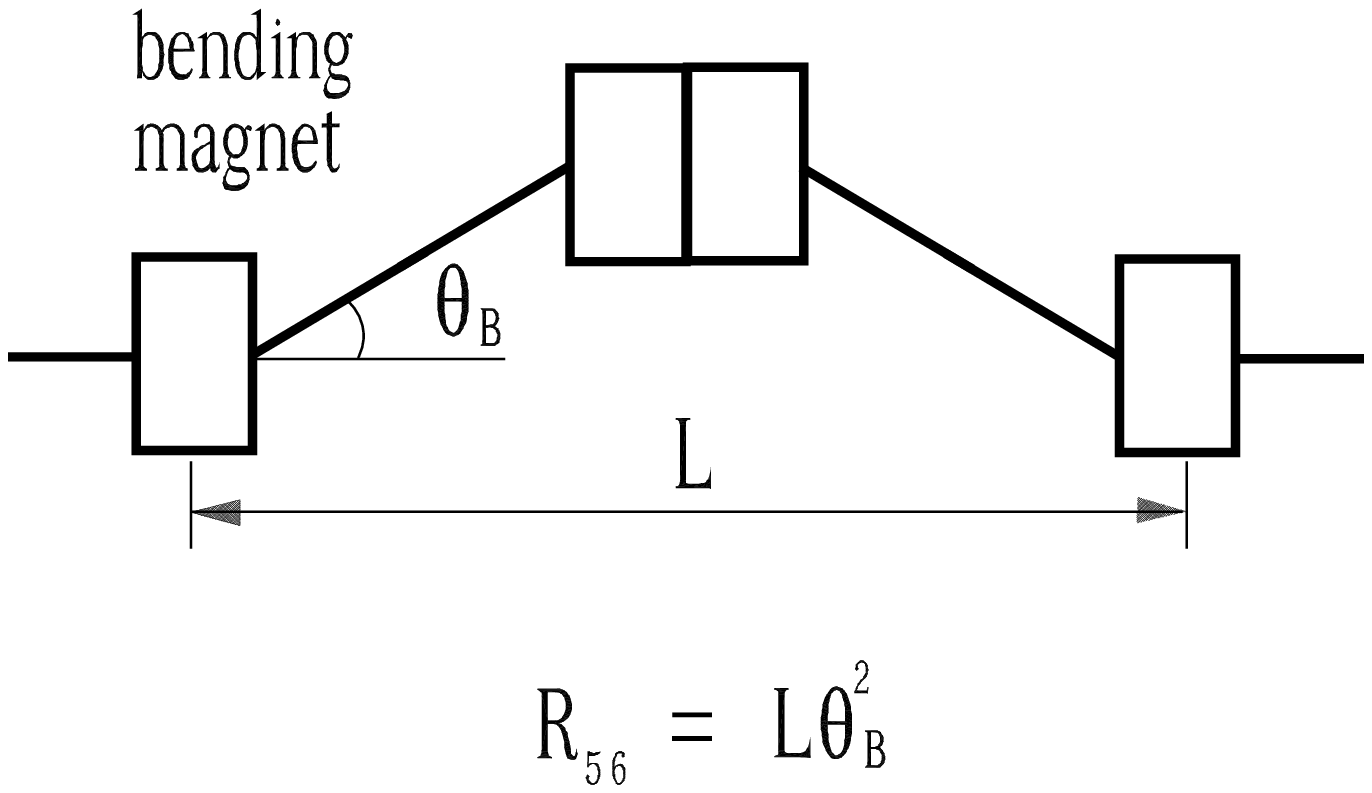,width=0.8\textwidth}
\end{center}
\caption{Schematic of dispersion section
}
\label{fig:ads}
\end{figure}

In two-undulator klystron, energy modulation occurs in the first
undulator and the conversion to density modulation in the dispersion
section. The dispersion section is designed to introduce the energy
dependence of a particle's path length, $\Delta z = R_{56}\delta{\cal
E}/{\cal E}_{0}$. Several designs are possible, but the simplicity of a
four-dipole magnet chicane is attractive because it adds no net
beamline bend angle or offset and allows simple tuning of the momentum
compaction, $R_{56}$, with a single power supply.
The trajectory of the electron
beam in the chicane has the shape of an isosceles triangle with
base length $L$.  The angle adjacent to the base,
$\theta_{\mathrm{B}}$, is considered to be small. For
ultra-relativistic electrons and small bend angles, the net $R_{56}$ of
the chicane is given by (see Fig.  \ref{fig:ads})

\begin{displaymath}
R_{56} = L\theta^{2}_{\mathrm{B}} \ .
\end{displaymath}

Optimal parameters of the dispersion section can be calculated in the
following way. The phase space distribution of the particles in the
first undulator is described in terms of the distribution function
$f(P,\psi_{0})$ written in "energy-phase" variables $P = {\cal E} -
{\cal E}_{0}$ and $\psi_{0} = 2\pi z^{\prime}/\lambda^{\prime} =
\omega(z/v-t)$, where ${\cal E}_{0}$ is the nominal energy of the
particle and $\omega$ is the angular frequency. Before entering the
first undulator, the electron energy distribution is assumed to be
Gaussian:

\begin{displaymath}
f_{0}(P) = \frac{1}{\sqrt{2\pi\langle(\Delta {\cal E})^{2}\rangle}}\exp
\left(- \frac{P^{2}}{2\langle(\Delta {\cal E})^{2}\rangle}\right) \ .
\end{displaymath}

\noindent The present study assumes the density modulation at the end
of first undulator to be very small and there is an energy modulation
$P_{0}\cos\psi_{0}$ only. Then the distribution function at the
entrance to the dispersion section is

\begin{displaymath}
f_{0}(P + P_{0}\sin\psi_{0}) \ .
\end{displaymath}

\noindent After passing through the dispersion section with dispersion
strength $\D\psi_{0}/\D P$, the electrons of phase $\psi_{0}$ and
energy deviation $P$ will come to a new phase $\psi_{0} +
P\D\psi_{0}/\D P$.  Hence, the distribution function becomes

\begin{displaymath}
f(P,\psi_{0}) = f_{0}\left(P + P_{0}\sin\left(\psi_{0} -
P\frac{\D\psi_{0}}{\D P}\right)\right).
\end{displaymath}

\noindent The dispersion strength parameter and compaction factor are
connected by the relation

\begin{displaymath}
\frac{\D\psi_{0}}{\D P} = \frac{2\pi}{\lambda^{\prime}}\frac{\D
z^{\prime}}{\D{\cal E}} = \frac{2\pi}{\lambda^{\prime}}
\frac{R_{56}}{{\cal E}_{0}} \ .
\end{displaymath}

\noindent The integration of the phase space distribution
over energy provides the beam density distribution, and the Fourier
expansion of this function gives the harmonic components of the density
modulation converted from the energy modulation \cite{k}. At the
dispersion section exit, we may express current $I$ in the form

\begin{eqnarray}
& \mbox{} &
I =
I_{0}\int\limits^{\infty}_{-\infty}f(P,\psi_{0})\D P =
I_{0} + 2I_{0}\sum^{\infty}_{n=1}\exp\left[-\frac{1}{2}
n^{2}\langle(\Delta {\cal E})^{2}\rangle\left(\frac{\D\psi_{0}}{\D
P}\right)^{2} \right]
\nonumber\\
& \mbox{} &
\times
J_{n}\left(nP_{0}\frac{\D\psi_{0}}{\D P}\right)\cos(n\psi_{0}) \ .
\label{eq:1}
\end{eqnarray}

\noindent We find a set of harmonic waves, of which the fundamental
term, with angular frequency $\omega$, is the one of
importance in an amplifier. This fundamental involves the phase
variation $\cos\psi$, but it is multiplied by the amplitude term

\begin{displaymath}
a_{\mathrm{out}} = 2J_{1}\left(P_{0}\frac{\D\psi_{0}}{\D P}\right)
\exp\left[-\frac{1}{2}
\langle(\Delta {\cal E})^{2}\rangle\left(\frac{\D\psi_{0}}{\D
P}\right)^{2} \right] \ .
\end{displaymath}

\noindent For small input signal we may assume that the argument of
the Bessel function is small. The function $J_{1}(X)$ approaches $X/2$
for small $X$; thus the microbunching approaches

\begin{displaymath}
a_{\mathrm{out}} = P_{0}\frac{\D\psi_{0}}{\D P}
\exp\left[-\frac{1}{2}
\langle(\Delta {\cal E})^{2}
\rangle\left(\frac{\D\psi_{0}}{\D P}\right)^{2}
\right] \ .
\end{displaymath}

\noindent We see that microbunching depends greatly on the choice of
the dispersion section strength. An attempt to increase of the
amplitude of the fundamental harmonic, by increasing the strength of
dispersion section, is countered by a decrease of
the exponential factor. The  microbunching $a_{\mathrm{out}}$  has
clearly a maximum

\begin{equation}
\left(a_{\mathrm{out}}\right)_{\mathrm{max}} = \frac{P_{0}}
{\sqrt{2.72\langle(\Delta {\cal E})^{2}\rangle}}
\label{eq:dm1}
\end{equation}

\noindent and the optimum strength of the dispersion section is

\begin{equation}
\left(\frac{\D\psi_{0}}{\D P}\right)_{\mathrm{max}} =
\frac{1}{\sqrt{\langle(\Delta {\cal E})^{2}\rangle}} \ .
\label{eq:dso}
\end{equation}

\noindent Consider the situation at $\sqrt{\langle(\Delta {\cal
E})^{2}\rangle} = 5 \ {\mathrm{keV}}$, ${\cal E}_{0} = 500 \
{\mathrm{MeV}}$, $\lambda = 30 \ {\mathrm{nm}}$. Appropriate
substitution in (\ref{eq:dso}) show that the optimal compaction factor
is equal to $R_{56} = L\theta^{2}_{\mathrm{B}} = 500 \
\mu{\mathrm{m}}$.

\section{The properties of a klystron amplifier}

We can now piece together
the information we have obtained in the preceding sections, and discuss
the behaviour of the klystron as an amplifier of the electron beam
intensity modulation. The present section is concerned with cascade
amplifiers designed for conditions approximating maximum gain per
cascade pass.

\subsection{Klystron amplifier gain}

In the section 2 we learned that when a modulated electron beam
passes through an undulator, the radiation field modulates the energy
of the electrons. In order to calculate the energy modulation we can
use (\ref{eq:r}) and (\ref{eq:m1}) and note that energy modulation at
the undulator exit generally can be written as $p\cos\psi_{0} +
q\sin\psi_{0}$, or as $\rho\cos(\psi_{0} + \delta)$.  When the gain per
cascade pass is high enough $(a_{\mathrm{out}} \gg a_{\mathrm{in}})$,
the cascades can be considered independent.  In this case we do not see
the effect of phase shift $\delta$, but see only total amplitude of
modulation equal to $\rho = \sqrt{p^{2}+q^{2}}$.

We are going to apply (\ref{eq:r}) to our analysis of energy modulation
taking advantage of similarity techniques  we used in section 2.1.  The
beam, of peak current $I_{0}$, is modulated in energy by the input
undulator, and acquires an energy modulation amplitude on the electron
beam axis, of amount

\begin{displaymath}
\frac{\delta{\cal E}}{{\cal E}} =
F(\beta)\left(\frac{\delta{\cal E}}{{\cal E}}\right)_{0} \ ,
\end{displaymath}

\noindent where

\begin{displaymath}
\left(\frac{\delta{\cal E}}{{\cal E}}\right)_{0} =
\frac{a_{\mathrm{in}}\theta^{2}_{\mathrm{w}}\omega L_{\mathrm{w}}}
{2c}\frac{I_{0}}{\gamma I_{\mathrm{A}}} \ ,
\end{displaymath}

\noindent and universal function $F(\beta)$ should be calculated
numerically by using (\ref{eq:r}). The physical implication of this
result are best understood by considering some limiting
cases. Equations (\ref{eq:me1}) and (\ref{eq:me2}) give us an idea
what we should expect in the case of wide and thin electron beam.  We
have asymptotically:

\begin{displaymath}
F(\beta) \to 1/(2\beta) \qquad {\mathrm{as}}\qquad \beta \to \infty \ ,
\end{displaymath}

\begin{displaymath}
F(\beta) \to \sqrt{
\pi^{2}/4 +
[\ln(1/\beta) - 1]^{2}} \qquad {\mathrm{as}}\qquad \beta \to 0 \ .
\end{displaymath}

\noindent A natural and interesting choice is to calculate the gain in
thin beam case which we can identify with maximal gain.
Note that the dependence of the factor $F(\beta)$ on the exact size of
the bunch in the thin beam asymptote is rather weak. Thus $F(\beta)
\simeq 1$ is a reasonable approximation. Finally, the energy modulation
amplitude can be estimated simply as:

\begin{equation}
\frac{\delta{\cal E}}{{\cal E}} \simeq
\left(\frac{\delta{\cal E}}{{\cal E}}\right)_{0}  =
\frac{a_{\mathrm{in}}\theta^{2}_{\mathrm{w}}\omega L_{\mathrm{w}}}
{2c}\frac{I_{0}}{\gamma I_{\mathrm{A}}} \qquad {\mathrm{for}}\qquad
\beta < 1 \ .
\label{eq:me3}
\end{equation}

\noindent The energy modulated beam then enters the dispersion section,
in which density modulation is developed.  From equations
(\ref{eq:dm1}) and (\ref{eq:me3}) the over-all  gain per
cascade pass at resonance may be readily be obtained if optimum
dispersion section strength is assumed; thus

\begin{equation}
G_{0} = \frac{a_{\mathrm{out}}}{a_{\mathrm{in}}} \simeq
\frac{\theta_{\mathrm{w}}^{2}\omega_{0}L_{\mathrm{w}}}
{2\sqrt{2.72}c}\frac{I_{0}}
{\Delta\gamma I_{\mathrm{A}}} \qquad {\mathrm{for}}\qquad \beta \ < \ 1
\ ,
\label{eq:g1}
\end{equation}

\noindent where the following notation has been introduced:
$\Delta\gamma = \sqrt{\langle(\Delta{\cal
E})^{2}\rangle/m_{\mathrm{e}}^{2}c^{4}}$.
Putting  $\theta_{\mathrm{w}} = K/\gamma$, $L_{\mathrm{w}}
= N_{\mathrm{w}}\lambda_{\mathrm{w}}$, $\omega_{0} = 2\pi/\lambda
= 4\pi\gamma^{2}_{z}c/\lambda_{\mathrm{w}}$ and $\gamma^{2}_{z} =
\gamma^{2}/(1+K^{2})$ in (\ref{eq:g1})  gives

\begin{equation}
G_{0} \simeq  1.2\pi
\frac{K^{2}}{(1+K^{2})}\frac{N_{\mathrm{w}}I_{0}}{\Delta\gamma
I_{\mathrm{A}}} \qquad {\mathrm{for}}\qquad \beta \ < \ 1 \ .
\label{eq:g2}
\end{equation}

\noindent The equation (\ref{eq:g2}) tells us the maximal
gain of cascade as a function of number of undulator periods
$N_{\mathrm{w}}$, undulator parameter $K$, peak current $I_{0}$ and rms
local energy spread $\Delta\gamma$.

For conventional XFELs the following problem exists. The
required very small transverse beam emittance can be obtained only in
a photoinjector.  The bunch length must be shortened.
The distribution of particles in phase space is given either by the
injector characteristics and the injection process.
Longitudinal phase
space can be exchanged by special application of magnetic and RF
fields. This is done in a specially designed beam transport line
consisting of a nonisochronous transport line (magnetic chicane) and an
accelerating section installed at the beginning of the bunch
compressor.  Liouville's theorem is not violated because the energy
spread in the beam has been increased through the phase dependent
acceleration in the bunch compression system.

The cascade gain with continually optimized
strength of the dispersion section would be directly proportional to
$I_{0}/ \Delta\gamma$.
We have a rather surprising result. We know that
ratio $I_{0}/ \Delta\gamma$ depends on the parameters of
photoinjector but not on the bunch compression. We expect this ratio to
depend only on the longitudinal emittance. So the simple result says
that the klystron gain is independent of absolute value of peak current.
The change in peak current with change of bunch length is
compensated by the larger energy spread. Since $I_{0}/\Delta\gamma$
is independent of bunch length, the gain would then be independent of
bunch compression.

The formula for the gain which we derived
(\ref{eq:g2}) refer to the case of the helical undulator.
For somewhat wider generality, although we are still making some
special assumption about  undulator magnetic structure, we shall
calculate the characteristics of the klystron amplifier with
a planar undulator. The magnetic field on the axis of the planar
undulator is given by $\vec{H} =
\vec{e}_{x}H_{\mathrm{w}}\cos\left(k_{\mathrm{w}}z\right)$.
The explicit expression for the electron velocity in the
field of the planar undulator has the form:
$v_{y} = -\vec{e}_{y}c\theta_{\mathrm{w}}\sin(k_{\mathrm{w}}z)$,
where $\theta_{\mathrm{w}} = K/\gamma = \lambda_{\mathrm{w}}
eH_{\mathrm{w}}/(2\pi m_{\mathrm{e}}c^{2}\gamma)$.
It is not hard to go through the derivation of electron beam energy
modulation again.  If we do that, and calculate the gain the same way,
we get

\begin{equation}
G_{0} = \frac{a_{\mathrm{out}}}{a_{\mathrm{in}}} \simeq
\frac{\theta_{\mathrm{w}}^{2}A^{2}_{\mathrm{JJ}}\omega_{0}L_{\mathrm{w}}}
{4\sqrt{2.72}c}\frac{I_{0}}
{\Delta\gamma I_{\mathrm{A}}}  \qquad {\mathrm{for}}\qquad \beta \ < \
1\ ,
\label{eq:g3}
\end{equation}

\noindent where

\begin{displaymath}
A_{\mathrm{JJ}} = [J_{0}(Q)-J_{1}(Q)] \ ,
\end{displaymath}

\noindent $J_{n}(Q)$ is a Bessel function of $n$th order,

\begin{displaymath}
Q = \theta^{2}_{\mathrm{w}}\omega_{0}/(8k_{\mathrm{w}}\gamma^{2})
= K^{2}/(4+2K^{2}) \ .
\end{displaymath}

\noindent When we simplified the expression for $Q$, we used the
resonance condition for the planar undulator $\omega_{0} =
2\gamma^{2}k_{\mathrm{w}} /[c(1+K^{2}/2)]$. It is convenient to rewrite
the expression (\ref{eq:g3}) in the form:

\begin{equation}
G_{0} \simeq  1.2\pi
\frac{A^{2}_{\mathrm{JJ}}K^{2}}{(2+K^{2})}
\frac{N_{\mathrm{w}}I_{0}}{\Delta\gamma I_{\mathrm{A}}}
\qquad {\mathrm{for}}\qquad \beta \ < \ 1 \ .
\label{eq:g4}
\end{equation}

The significance of the proposed scheme cannot be fully appreciated
until we determine typical values of the gain per cascade pass that
can be expected in practice. Let us present a specific numerical
example for the case of a klystron amplifier with a planar undulator.
With the numerical values $\lambda_{\mathrm{w}} = 3 \ {\mathrm{cm}}$,
$K = 1.42$, $\gamma = 10^{3}$, the resonance value of wavelength is
$\lambda = 30 \ {\mathrm{nm}}$. If the number of the undulator period
is $N_{\mathrm{w}} = 100$, normalized transverse emittance
$\epsilon_{\mathrm{n}} = 2 \pi \ \mu{\mathrm{m}}$, and betatron
function is equal to the undulator length, the diffraction
parameter is about $\beta \simeq 0.4$. Remembering the results of
section 2, we come to the conclusion that we can treat this situation
as an klystron amplifier with thin electron beam. It is shown that the
residual energy spread in the TTF FEL injector is on the order of a few
keV only, even at a peak current of 100 A. If $I_{0} = 100 \
{\mathrm{A}}$, and $\sqrt{\langle(\Delta {\cal E})^{2}\rangle} = 5 \
{\mathrm{keV}}$, appropriate substitution in (\ref{eq:g4}) shows that
the intensity gain per cascade pass is about $G^{2}_{0} \simeq 10^{4}$.

It has been seen above that there are definite upper limits to the gain
of a two undulator klystron with a given current.  It is
obvious that a gain much higher than that allowed by these limits may
be obtained by using more than one stage of amplification. In the
three-undulator type of klystron, called the cascade-amplifier
klystron, the additional undulator (and dispersion section), which lies
between the input and output undulators, has no need of radiation
phase matching. Second cascade is excited by the bunched beam that
emerges from the first cascade, and it produces further bunching of the
beam.  Cascade bunching in a high-gain cascade amplifier results in
current components just like those of simple bunching; the equivalent
intensity gain, which determines the cascade bunching, is given by
$G^{2}_{\mathrm{eq}} = G_{0}^{2m}$, where $m$ is the number of
cascades.  In practice, with a two-cascade klystron amplifier, an
intensity gain in excess of 80 dB may be obtained at wavelengths around
30 nm. This simple description is valid when beam modulation amplitude
at the exit of the last cascade is much smaller than unity.  What
happens when this condition is not met is discussed later.

Fluctuations of the electron beam current serve as an input signal to
the XFEL. The initial modulation of the electron beam is defined by the
shot nose and has a white spectrum. When the electron beam enters the
first undulator, the presence of the beam modulation at frequencies
close to resonance frequency initiates the process of radiation.
The study has shown (see section 5) that the mean square value of
input is about $\langle a_{\mathrm{in}}^{2}\rangle \simeq
3\cdot 10^{-7}$ at $I_{0} = 100 \ {\mathrm{A}}$, $N_{\mathrm{w}} = 100$
and $\lambda = 30 \ {\mathrm{nm}}$.  It is seen that saturation in  the
cascade klystron should occur at an exit of the second cascade in this
situation.  Thus, we conclude that a klystron amplifier which consists
of a succession of 2-3 cascades can operate as a soft X-ray SASE FEL.

A disadvantage of using conventional FEL amplifier at small peak
current is the enormously  long undulator that is required.
Let us calculate the parameters of the conventional
SASE FEL for the value $I_{0} = 100 \ {\mathrm{A}}$ of the peak
current.  Suppose $\sqrt{\langle(\Delta {\cal E})^{2}\rangle} = 5 \
{\mathrm{keV}}$, $\epsilon_{\mathrm{n}} = 2\pi \mu{\mathrm{m}}$,
$\gamma = 1000$, $\lambda_{\mathrm{w}} = 3 \ {\mathrm{cm}}$, $K =
1.42$, focusing beta function is equal to 3 m. Numerical solution of
the corresponding 3-D eigenvalue equation (see \cite{book}) demonstrates
that the field gain length should be about $l_{\mathrm{g}} \simeq 3.7 \
{\mathrm{m}}$.  Well-known that saturation in a SASE FEL with uniform
undulator occurs after approximately 10 exponential field gain lengths.
Remembering the total undulator length
of the two-cascade klystron amplifier, we
come to the conclusion that a significant impruvement  in this
parameter for klystron amplifier can be achieved.

\subsection{Klystron amplifier bandwidth}

All of the foregoing discussion of klystron amplifier gain has been
concerned solely with the gain at resonance - that
is $\omega = \omega_{0}$. Nothing has been said about amplification
bandwidth. Now, we would like to find out how the gain varies in the
circumstance that the seed signal frequency $\omega$ is nearly, but not
exactly, equal to $4\pi\gamma^{2}_{z} c/\lambda_{\mathrm{w}}$.
It is not hard to go through the derivation of electron beam energy
modulation again. If we take $C \ne 0$, the solution of the wave
equation (\ref{eq:u8}) has the form

\begin{eqnarray}
& \mbox{} &
\tilde{E}(z,\vec{r}_{\perp}) = \frac{\I
e\theta_{\mathrm{w}}\omega a_{\mathrm{in}}}{2c}
\int\limits^{z}_{0}\frac{\D z^{\prime}}{z -
z^{\prime}}\exp(-\I C z^{\prime})
\nonumber\\
& \mbox{} &
\times
\int\D\vec{r}^{\prime}_{\perp}n_{0}(\vec{r}_{\perp}^{\prime})
\exp\left[\frac{\I\omega\mid \vec{r}_{\perp} -
\vec{r}^{\prime}_{\perp}\mid^{2}}{2c(z- z^{\prime})}\right] \ .
\label{eq:r21}
\end{eqnarray}

\noindent Using (\ref{eq:m1}) the energy modulation achieved at nonzero
detuning parameter can be written as

\begin{eqnarray}
& \mbox{} &
\delta{\cal E} = - \I\exp(\I\psi_{0}) \frac{
e^{2}\theta^{2}_{\mathrm{w}}\omega a_{\mathrm{in}}}{4c}
\int\limits^{z}_{0}\D z^{\prime}\exp(\I C z^{\prime})
\int\limits^{z^{\prime}}_{0}\frac{\D z^{\prime\prime}}{z^{\prime} -
z^{\prime\prime}}\exp(-\I C z^{\prime\prime})
\nonumber\\
& \mbox{} &
\times
\int\D\vec{r}^{\prime}_{\perp}n_{0}(\vec{r}_{\perp}^{\prime})
\exp\left[\frac{\I\omega\mid \vec{r}_{\perp} -
\vec{r}^{\prime}_{\perp}\mid^{2}}{2c(z^{\prime}-
z^{\prime\prime})}\right] + {\mathrm{C.C.}}
\label{eq:me21}
\end{eqnarray}

\noindent Now we shall calculate the energy modulation on the electron
beam axis. When $r = 0$ and beam profile is Gaussian, we have

\begin{displaymath}
\frac{\delta{\cal E}}{{\cal E}} = - \I e^{\I\psi_{0}} \frac{
a_{\mathrm{in}}\theta^{2}_{\mathrm{w}}\omega}{4c}\frac{I_{0}}{\gamma
I_{\mathrm{A}}}\int\limits^{z}_{0}\D z^{\prime}\int\limits^{z^{\prime}}_{0}\D
z^{\prime\prime}\frac{e^{\I C(z^{\prime}- z^{\prime\prime})}}
{z^{\prime} - z^{\prime\prime} + \I k\sigma^{2}} + {\mathrm{C.C.}}
\end{displaymath}

\noindent Let us see what happens if the diffraction parameter $\beta$
is large. In this case we have asymptotically:

\begin{displaymath}
\frac{\delta{\cal E}}{{\cal E}} = - e^{\I\psi_{0}} \frac{
a_{\mathrm{in}}\theta^{2}_{\mathrm{w}}}{4\sigma^{2}}
\frac{I_{0}}{\gamma I_{\mathrm{A}}}
\int\limits^{z}_{0}\D z^{\prime}\int\limits^{z^{\prime}}_{0}
\D z^{\prime\prime}e^{\I C(z^{\prime}-z^{\prime\prime})} +
{\mathrm{C.C.}}
\end{displaymath}

\noindent The integrals over $z^{\prime}$ and $z^{\prime\prime}$ in the
latter equation are calculated analytically

\begin{eqnarray}
& \mbox{} &
e^{\I\psi_{0}}\int\limits^{z}_{0}\D
z^{\prime}\int\limits^{z^{\prime}}_{0} \D z^{\prime\prime}e^{\I
C(z^{\prime}-z^{\prime\prime})} + {\mathrm{C.C.}}
\nonumber\\
& \mbox{} &
=
\frac{\sin^{2}(Cz/2)}{(C/2)^{2}}\cos\psi_{0} +
\frac{(Cz/2) - \sin(Cz/2)\cos(Cz/2)}{(C/2)^{2}}\sin\psi_{0} \ .
\label{eq:int1}
\end{eqnarray}

\noindent Taking into account the definition of the detuning parameter
we find that $C$ is connected by a simple
relation with the frequency deviation: $\omega - \omega_{0} =
\Delta\omega = 2\gamma^{2}_{z}C$.  If we follow through the simple
algebra we find that

\begin{eqnarray}
& \mbox{} &
\frac{\delta{\cal E}}{{\cal E}} = - \frac{
a_{\mathrm{in}}\theta^{2}_{\mathrm{w}}}{4}\frac{L^{2}_{\mathrm{w}}}
{\sigma^{2}} \frac{I_{0}}{\gamma I_{\mathrm{A}}}\left\{
\frac{\sin^{2}[\pi N_{\mathrm{w}}\Delta\omega/\omega_{0}]} {[\pi
N_{\mathrm{w}}\Delta\omega/\omega_{0}]^{2}}\cos\psi_{0}
\right.
\nonumber\\
& \mbox{} &
\left.
+
\frac{[\pi N_{\mathrm{w}}\Delta\omega/\omega_{0}] -
\sin[\pi N_{\mathrm{w}}\Delta\omega/\omega_{0}]
\cos[\pi N_{\mathrm{w}}\Delta\omega/\omega_{0}]}
{[\pi N_{\mathrm{w}}\Delta\omega/\omega_{0}]^{2}}
\sin\psi_{0} \right\} \ .
\label{eq:int2}
\end{eqnarray}

\begin{figure}[tb]
\begin{center}
\epsfig{file=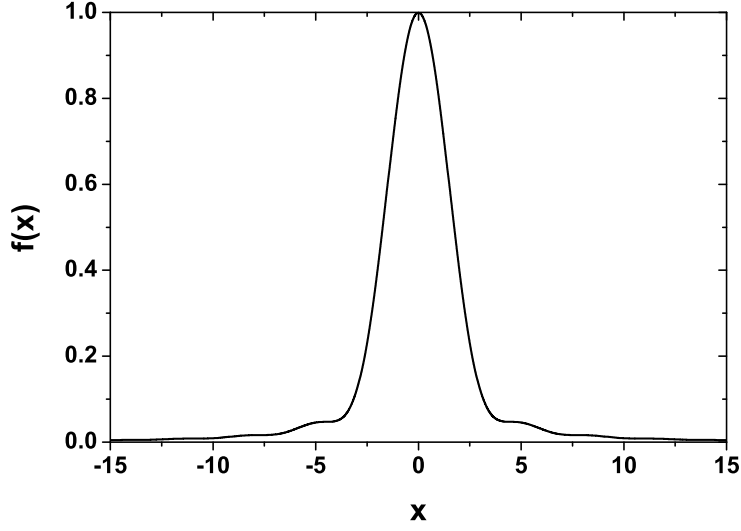,width=0.8\textwidth}
\end{center}
\caption{Frequency response curve}
\label{fig:fx}
\end{figure}

\noindent The energy modulation at the undulator exit can be written as
$p\cos\psi_{0} + q\sin\psi_{0}$. The intensity gain
$G^{2} = a^{2}_{\mathrm{out}}/a^{2}_{\mathrm{in}}$ is proportional to
$(p^{2}+q^{2})$. It is interesting to plot this gain as a function of
input signal frequency in order to see how sensitive it is to
frequencies near the resonant frequency $\omega_{0}$. We show such a
plot in Fig.  \ref{fig:fx}, where the
following notation has been introduced:  $x = \pi
N_{\mathrm{w}}\Delta\omega/\omega_{0}$,

\begin{equation}
\left[\frac{G(x)}{G(0)}\right]^{2} = f(x) = \frac{\sin^{4}(x)}{x^{4}} +
\frac{[x-\sin(x)\cos(x)]^{2}}{x^{4}} \ .
\label{eq:df}
\end{equation}

\noindent  The curve falls rather abruptly to zero for
$(\omega-\omega_{0}) = 1.7\omega_{0}/(\pi N_{\mathrm{w}})$ and never
regains significant size for large frequency deviations. So, for large
value of diffraction parameter, the frequency response curve is $f(x)$.
This is also true for the practically important case in which the
diffraction parameter is close to unity. Let's examine the implication
of this results for a real klystron amplifier. Suppose that the
electron beam is in the undulator for a reasonable length, say for 100
periods.  Then we can calculate that the FWHM of the amplification
bandwidth is equal to $(\Delta\omega)_{\mathrm{FWHM}}/\omega_{0} =
1.1/N_{\mathrm{w}} \simeq 1 \ \%$.

We have, then, a clear picture of the nature of the operation of
the klystron amplifier with nonzero detuning parameter, and a
calculation of the bandwidth to be obtained from it. A detail
analytical theory of the amplification at $C \ne 0$ and finite
transverse electron beam size is rather complicated, but the discussion
which we have given here is enough for ordinary purposes.

\subsection{Output of klystron amplifier}

The amplification process in the klystron amplifier can be divided into
two stages, linear and nonlinear. During the linear stage of
amplification, significant growth of the beam modulation amplitude and
of the electromagnetic field amplitude takes place. Nevertheless, the
beam modulation is much less than unity in the cascade undulators, and
the largest fraction of the radiation power is produced in the output
undulator, when the beam modulation becomes about unity.

Suppose we have a modulated electron beam at the exit of the last
cascade.  The problem of electromagnetic wave radiation in the output
undulator refers to a class of self-consistent problem. There is
considerable analogy between the nonlinear mode operation of the
conventional FEL amplifier and klystron amplifier. In the high
gain linear regime of a conventional FEL amplifier the typical length
of the change in the radiation field amplitude is about the gain length
$l_{\mathrm{g}} \simeq \Gamma^{-1}$, where for the one dimensional
model

\begin{displaymath}
\Gamma = \left[\frac{I_{0}\theta^{2}_{\mathrm{w}}\omega}
{2c\gamma^{2}_{z}\gamma I_{\mathrm{A}}\sigma^{2}}\right]^{1/3} \ .
\end{displaymath}

\noindent In the case of a uniform output undulator, the bunched beam
effectively interacts with the electromagnetic wave along a length
which is of the order of the gain length $l_{\mathrm{g}}$. At this
stage of the self-interaction electrons lose the visual fraction of
their energy which results in the violation of the resonance condition.
As a result, the beam is overmodulated, most electrons fall into the
accelerating phase of the ponderomotive potential and the electron beam
starts to absorb power from the electromagnetic wave.
This is analogous to the situation with a conventional FEL amplifier.
Like the efficiency of the conventional FEL amplifier, the efficiency
$\eta$ of the klystron amplifier with a long enough uniform output
undulator is limited by the value of the efficiency parameter $\rho$:

\begin{displaymath}
\eta \simeq \rho = c\gamma_{z}^{2}\Gamma/\omega \ ,
\end{displaymath}

\noindent It should be note that the saturation efficiency of the
klystron amplifier does not depend on the amplitude of the input signal
when the klystron amplifier operates in the high-gain regime, i.e. when
$a_{\mathrm{in}}/a_{\mathrm{out}} \ll 1$.

For some purposes it is convenient to express $\rho$
in a different form

\begin{equation}
\rho = \left[\frac{c\gamma_{z}}{2\omega\sigma}\right]^{2/3}
\left[\frac{I_{0}}{\gamma I_{\mathrm{A}}}\right]^{1/3}
\left[\frac{K^{2}}{1+K^{2}}\right]^{1/3} \ .
\label{eq:rho}
\end{equation}

\noindent Thus the efficiency of the klystron amplifier, defined in
this particular way, depends on only the one-third power of peak
current, a very weak dependence. Let us make a calculation of $\eta$
for some cases. Suppose $I_{0} = 100 \ {\mathrm{A}}$,
$\epsilon_{\mathrm{n}} = 2\pi \mu{\mathrm{m}}$, $\gamma = 1000$,
$\lambda_{\mathrm{w}} = 3 \ {\mathrm{cm}}$, $K = 1.42$,
focusing beta function is equal to 3 m; then by equation
(\ref{eq:rho}) it follows that $\eta \simeq 0.13\ \%$. Returning to
(\ref{eq:rho}), we could imaging the injector linac running at a $I_{0}
= 2.5 \ {\mathrm{kA}}$. As we mentioned above, the dependence of $\eta$
on the peak current is rather weak. As a result, the efficiency is
increased by only a factor three  ($\eta \simeq 0.4 \ \%$) for  $I_{0}
= 2.5 \ {\mathrm{kA}}$.

Let us present a specific numerical example for the case of a
fourth-generation light source. One of the key parameters to compare
different radiation sources is their brilliance, which is simply given
by the spectral flux divided by the transverse photon phase space.
We illustrate with numerical example the potential of the proposed
klystron amplifier scheme for the SASE FEL at the TESLA Test Facility
accelerator. The average brilliance of the conventional TTF SASE FEL
at a wavelength of $\lambda \simeq 30 \ {\mathrm{nm}}$ is about

\begin{displaymath}
B_{\mathrm{av}} \simeq
10^{22} \ {\mathrm{photons/s/0.1\% BW/mm}}^{2}/{\mathrm{mrad}}^{2}
\end{displaymath}

\noindent for the case when $I_{0} = 2.5 \ {\mathrm{kA}}$ \cite{x1}.
Average photon flux and average brilliance varies simply as efficiency
$\eta$. Since $B_{\mathrm{av}} \propto \eta$, the dependence of
$B_{\mathrm{av}}$ on the electron peak current is contained completely
in the term $\rho$. If $I_{0} = 100 \ {\mathrm{A}}$, appropriate
substitution in (\ref{eq:rho}) shows that the average brilliance of the
cascade klystron amplifier without bunch compression at a wavelength of
$\lambda \simeq 30 \ {\mathrm{nm}}$ is about

\begin{displaymath}
B_{\mathrm{av}} \simeq 3\cdot 10^{21}
{\mathrm{photons/s/0.1\%BW/mm}}^{2}/{\mathrm{mrad}}^{2} \ .
\end{displaymath}

\noindent On the other hand, at a wavelength of 30 nm
brilliance of a third-generation synchrotron light source  is equal to

\begin{displaymath}
B_{\mathrm{av}} \simeq 10^{17}
{\mathrm{photons/s/0.1\%BW/mm}}^{2}/{\mathrm{mrad}}^{2} \ .
\end{displaymath}

\noindent A comparison to synchrotron light sources
shows the drastic improvement of average brilliance at the klystron
sources. The average brilliance of the klystron facility operating
without bunch compression in the injector linac surpasses the
spontaneous undulator radiation from third-generation synchrotron
radiation facilities by four or more orders of magnitude.
However, it is not only the high average brilliance which makes
a klystron amplifier facility a unique research tool in the VUV and
soft X-ray regime, the radiation from this facility shows a very high
peak brilliance too.  Average brilliance is normalized to seconds at
the highest possible repetition rate while the peak brilliance is the
brilliance scaled to the length of a single pulse. Another type of
current dependence are expected for peak brilliance, a 4/3 power
dependence $B_{\mathrm{peak}} \propto I_{0}^{4/3}$. The peak
brilliance of the conventional TTF SASE FEL at a wavelength of $\lambda
\simeq 30 \ {\mathrm{nm}}$ is about

\begin{displaymath}
B_{\mathrm{peak}} \simeq
10^{29} \ {\mathrm{photons/s/0.1\% BW/mm}}^{2}/{\mathrm{mrad}}^{2} \ .
\end{displaymath}

\noindent On the other hand, the peak brilliance of the
cascade klystron amplifier without bunch compression at a wavelength of
$\lambda \simeq 30 \ {\mathrm{nm}}$ is about

\begin{displaymath}
B_{\mathrm{peak}} \simeq 10^{27}
{\mathrm{photons/s/0.1\%BW/mm}}^{2}/{\mathrm{mrad}}^{2} \ .
\end{displaymath}

\noindent  Decreasing the peak current also decreases the peak
brilliance of the SASE FEL radiation by about of factor 100, but this
is still 6 orders of magnitude higher than that of 3rd generation
synchrotron radiation sources.

A cascade klystron amplifier can be modified to increase significantly
the spectral brightness of output radiation. A reliable method to
increase the klystron amplifier efficiency consists in an adiabatic
change of the output undulator parameters (or, in other words, by the
use of so-called undulator tapering). When the electron bunch passes
through the last dispersion section the energy modulation leads to
an effective (nonlinear) compression of the particles.  Then the
bunched electron beam enters a tapered output undulator, and from the
very beginning produces strong radiation because of the large spatial
bunching. The strong radiation field produces a ponderomotive well
which is deep enough to trap the particles, since the original beam is
relatively cold. The radiation produced by these captured particles
increases the depth of the ponderomotive well, and they are effectively
decelerated.  As a result, much higher power can be achieved than for
the case of a uniform undulator.

In \cite{fd} we have described  the tapering undulator scheme for a
SASE FEL. Simulations using the code FAST provide a "full physics"
description of the process.  Despite the original spiking seeding
the process of the electron density modulation, we effectively trap a
significant fraction of the particles, and can achieve much higher
power than for the case of an untapered output undulator. Another
important feature of the radiation from a tapered undulator is the
significant suppression of the sideband growth in the nonlinear regime.
This means that in the proposed scheme the spectral brightness of the
radiation is increased proportionally to the radiation power. In the
case of a uniform undulator the peak brightness is reached at the
saturation point and is then reduced due to sideband growth.

\section{Klystron amplifier start-up from shot noise}

Till now we have considered the electron beam as a continuous medium
when describing the theory of klystron amplifier. To some extent this
is idealization, since in reality the beam current is produced by a
large number of moving electrons. If we consider the microstructure of
the electron current, we find that electrons enter the undulator
randomly in time and space. So, we can expect that the FEL amplifier
should possess intrinsic noise properties. Here it is relevant to
remember that FELs form a separate class of vacuum-tube devices. The
analysis of the noise properties of traditional vacuum-tube amplifiers
has always been an important problem. This has been mainly connected
with the practical need for reducing the intrinsic noise of the
amplifier. The result of the experience obtained during these
investigations shows that there always exists the fundamental effect of
shot noise originating from the random emission of the electrons from
the cathode. When we analyze this effect for the parameters of
traditional microwave amplifiers, we find that it is complicated. In
particular, suppression of the shot noise in some frequency band can
occur due to space charge effects. Besides the shot noise effect, there
are a number of different sources of noise which influence the
operation of traditional vacuum-tube amplifiers.

Fluctuations of the electron beam current serve as the input signal in
the XFEL. As for the FEL amplifier operating in short wavelength range,
its noise properties are defined only by the shot noise.  At optical
frequencies the quieting effect of space-charge limitation seems
negligible. An FEL amplifier which starts up from shot noise is
frequently known as a self-amplified spontaneous emission (SASE) FEL.
However, it is worth mentioning that such an essentially quantum
terminology does not reflect the actual physics of the process.  The
amplification process in the SASE FEL has its origin in the density
fluctuations in the electron beam. The latter effect is completely
classical.

The shot noise is the pure
fluctuations in number which correspond to the fact that photoemission
is a random process.  The rate at which electrons are emitted from a
photocathode is not constant in time. The emission process is a random
one and it is impossible to predict the time dependence of the
instantaneous current.  Any such random fluctuations in the beam
current correspond to an intensity modulation of the beam current at
all frequencies simultaneously - including, of course, the frequency to
which the undulator is tuned.  When the electron beam enters the first
undulator, the presence of the beam modulation at frequencies close to
the resonance frequency initiates the process of radiation.

The emission of electrons from the cathode is believed to be a Poisson
process, and from this assumption alone the total fluctuation in
current can be deduced.  Why is this the right rule, what is the
fundamental reason for it, and how is it connected to anything else?
The explanation in deep down in quantum mechanics.  When
electromagnetic fields are incident on a photosurface, a complex set of
events can transpire. The major steps in this process can be identified
as absorption of a quantum light energy (i.e. a photon) and the
transfer of that energy to an excited electron, transport of the
excited electron to the surface, and finally, release of the electron
from the surface. When laser light having a deterministic variation of
intensity over space and time is incident on a photosurface, the
fluctuations of the photons obey Poisson
statistics.  This explains the relation between fundamental
photon "shot noise" and shot noise in the electron beam.

When describing the physical principles, it is always important to find
a model which provides the possibility of an analytical description
without loss of essential information about the features of the random
process. In the first analysis of the problem, we adopt some rather
simplifying assumptions that are only occasionally met in practice. We
will investigate the klystron amplifier start-up from shot noise in the
framework of the one-dimensional model which assumes the input shot
noise and output radiation have a full transverse coherence. This
assumption allow us to assume that the input shot noise signal is
defined by the value of total beam current. In reality the fluctuations
of the electron beam current density are uncorrelated in the transverse
dimension. Using the notion of the beam radiated modes, we can say that
many transverse radiation modes are radiated when the electron beam
enters the undulator. The one-dimensional model can be used for the
calculations of statistical properties of transversely coherent output
current.  In practice such an assumption is valid for the current at
the exit of klystron with a thin electron beam.  During the
amplification process in a klystron, the number of transverse modes
decreases, and the contribution of the coherent radiation to the
total radiation power is increased up to full coherence. With this
assumption, attention can be concentrated completely on temporal
coherence effects.

We study the case when the initial modulation of the electron beam is
defined by the shot noise and has a white spectrum. Since in the linear
regime all the harmonics are amplified independently, we can use the
results of the steady-state theory for each harmonic and calculated the
corresponding Fourier harmonics of the output current. The electric
current in the time domain, $I(t)$, and its Fourier
transform, $\bar{I}(\omega)$, are connected

\begin{displaymath}
I(t) = \frac{1}{2\pi}\int\limits^{\infty}_{-\infty}
\bar{I}(\omega)\exp(-\I\omega t)\D\omega \ .
\end{displaymath}

\noindent The klystron amplifier has a response which can be
economically expressed as a frequency response curve $G(\omega)$. Thus,
the Fourier harmonic of the current at the dispersion section
exit and the Fourier harmonic of the input current are connected by the
relation:

\begin{equation}
\bar{I}_{\mathrm{out}}(\omega) = G(\omega)\bar{I}(\omega)
\qquad \omega > 0 \ .
\label{eq:s7}
\end{equation}

\noindent When
$\omega < 0$ the Fourier harmonic is defined by the relation
$\bar{I}^{*}(\omega) = \bar{I}(-\omega)$. It is convenient to isolate
explicitly the slowly varying complex amplitude

\begin{equation}
I(t) - I_{0} = \tilde{I}(t)e^{-\I\omega_{0}t}  +
{\mathrm{C.C.}} \ ,
\label{eq:sn1}
\end{equation}

\noindent where

\begin{displaymath}
\tilde{I}(t)e^{-\I\omega_{0}t} = \frac{1}{2\pi}
\int\limits^{\infty}_{0}\bar{I}(\omega)e^{-\I\omega t}\D\omega \ .
\end{displaymath}

The real output current consists of a carrier modulation of frequency
$\omega_{0}$ subjected to random amplitude and phase modulation. The
Fourier decomposition of the output current contains frequencies spread
about $\omega_{0}$. It is not possible in practice to resolve the
oscillations in $I(t)$ that occur at frequency of the carrier
modulation. A good experimental resolving time is of order of
picoseconds, at least four orders of magnitude too long to detect
oscillations at the frequency $\omega_{0}$. It is therefore appropriate
for comparison with experiment to average the theoretical results over
a cycle of oscillation of the carrier modulation. The deviation
$\delta I = I(t)-I_{0}$ has a zero cycle-average. According to
(\ref{eq:sn1}), the cycle average of the square of the current
deviation is

\begin{equation}
(\delta I)^{2} = 2\mid\tilde{I}(t)\mid^{2} \ .
\label{eq:d1}
\end{equation}

Shot noise is a random statistical process and statements about shot
noise are probability statements. These are most easily handled using
the concept of a statistical ensemble, drawn from statistical
mechanics. If the physical construction of the injector and set of the
external parameters (e.g supply voltage, photoinjector laser beam
parameters) which specify its state are known, then these
parameters define a definite statistical ensemble to which the injector
belongs.This ensemble consists of identical copies of the injector
subject to identical macroscopic external conditions. Any external
property of the injector, e.g. its laser light on the photosurface,
will depend, however, not only on these given parameters but also on
the exact microscopic internal state of the photocathode, and each
copy of the injector in the ensemble will lead to a different output.
We shall always denote ensemble averages by $\langle\cdots\rangle$.
In practice the XFEL injector is required to produce a single electron
bunch at a repetition rate of 100 - 100 000 bunch/s and
averaging symbol $\langle\cdots\rangle$ simply means the ensemble
average over electron bunches.

\subsection{Shot noise in a linear amplifier}

The output fluctuations, expressed as current
fluctuations, have a mean square value

\begin{displaymath}
\langle (\delta I_{\mathrm{out}})^{2}\rangle =
\frac{1}{2\pi^{2}}\int\limits^{\infty}_{0}\D\omega
\int\limits^{\infty}_{0}\D\omega^{\prime}G(\omega)G^{*}(\omega^{\prime})
\langle\bar{I}(\omega)\bar{I}^{*}(\omega^{\prime})\rangle \ .
\end{displaymath}

\noindent To calculate correlation of the spectral components of the
output beam current, we should consider a microscopic picture of the
electron beam current at the undulator entrance. We start the analysis
for the case of a rectangular electron pulse of finite duration $T$ and
then we go over to the limit of an infinitely long pulse. The electron
beam current is made up of moving electrons randomly arriving at the
entrance of the undulator:

\begin{displaymath}
I(t) = (-e)\sum^{N}_{k=1}\delta(t-t_{k}) \ ,
\end{displaymath}

\noindent where $\delta(\cdots)$ is the delta function, $(-e)$ is the
charge of the electron, $N$ is the number of electrons in a bunch and
$t_{k}$ is the random arrival time of the electron at the undulator
entrance. The beam current averaged over an ensemble of the bunches can
be written in the form:

\begin{displaymath}
\langle I(t)\rangle = \frac{(-e)N}{T} = - I_{0} \qquad {\mathrm{for}}
\quad - \frac{T}{2} < t < \frac{T}{2} \ .
\end{displaymath}

\noindent The electron beam current
$I(t)$ and its Fourier transform $\bar{I}(\omega)$ are connected by:

\begin{displaymath}
\bar{I}(\omega) = \int\limits^{\infty}_{-\infty}
e^{\I\omega t}I(t)\D t = (-e)\sum^{N}_{k=1}e^{\I\omega t_{k}} \ ,
\end{displaymath}

\begin{displaymath}
I(t) = \frac{1}{2\pi}\int\limits^{\infty}_{-\infty}
\bar{I}(\omega)e^{-\I\omega t}\D\omega =
(-e)\sum^{N}_{k=1}\delta(t-t_{k}) \ .
\end{displaymath}

Now we can calculate the first-order correlation of the complex Fourier
harmonics $\bar{I}(\omega)$ and $\bar{I}(\omega^{\prime})$:

\begin{displaymath}
\langle\bar{I}(\omega)\bar{I}(\omega^{\prime})\rangle
= e^{2}\left\langle\sum^{N}_{k=1}\sum^{N}_{n=1}\exp
(\I\omega t_{k} - \I\omega^{\prime}t_{n})\right\rangle \ .
\end{displaymath}

\noindent Expanding this relation, we can write:

\begin{eqnarray}
& \mbox{} &
\langle\bar{I}(\omega)\bar{I}(\omega^{\prime})\rangle
\nonumber\\
& \mbox{} &
=
e^{2}\left\langle\sum^{N}_{k=1}\exp
[\I(\omega - \omega^{\prime})t_{k}]\right\rangle
+
e^{2}\left\langle\sum_{k\ne n}\exp
(\I\omega t_{k} - \I\omega^{\prime}t_{n})\right\rangle
\nonumber\\
& \mbox{} &
=
e^{2}\sum^{N}_{k=1}\langle\exp
[\I(\omega - \omega^{\prime})t_{k}]\rangle
+
e^{2}\sum_{k\ne n}\langle\exp
(\I\omega t_{k})\rangle\langle\exp(-
\I\omega^{\prime}t_{n})\rangle \ .
\label{eq:s1}
\end{eqnarray}

\noindent The probability of the arrival of an electron during a time
interval $(t, \ t+\D t)$ is equal to $\D t/T$.
It is easy to find that

\begin{equation}
\langle\exp(\I\omega t_{k})\rangle
= \frac{1}{T}\int\limits^{T/2}_{-T/2}e^{\I\omega t_{k}}\D t_{k}
= \bar{F}(\omega) \ ,
\label{eq:s2}
\end{equation}

\noindent where

\begin{displaymath}
\bar{F}(\omega) = \left(\frac{\omega T}{2}\right)^{-1}
\sin\left(\frac{\omega T}{2}\right) \ .
\end{displaymath}

\noindent Substituting  (\ref{eq:s2}) to the (\ref{eq:s1}), we obtain:

\begin{equation}
\langle\bar{I}(\omega)\bar{I}^{*}(\omega^{\prime})\rangle
= e^{2}N\bar{F}(\omega-\omega^{\prime})
+ e^{2}N(N-1)\bar{F}(\omega)\bar{F}^{*}(\omega^{\prime}) \ .
\label{eq:s3}
\end{equation}

\noindent When

\begin{equation}
N\mid\bar{F}(\omega)\mid^{2} \ll 1 \ ,
\label{eq:s8}
\end{equation}

\noindent we can write the following expression for the first-order
spectral correlation:

\begin{equation}
\langle\bar{I}(\omega)\bar{I}^{*}(\omega^{\prime})\rangle
= e^{2}N\bar{F}(\omega-\omega^{\prime})  \ .
\label{eq:s9}
\end{equation}

\noindent Let us consider the case of an infinitely long electron
pulse with the homogeneous linear density and current $I_{0}$.  In this
limit $(T \to \infty,\ N \to \infty, \ eN/T = {\mathrm{const.}})$  the
first-order correlation of the complex Fourier harmonics has the form:

\begin{equation}
\langle\bar{I}(\omega)\bar{I}^{*}(\omega^{\prime})\rangle
= eI_{0}\lim_{T\to\infty}\left[T\bar{F}(\omega-\omega^{\prime})
\right] = 2\pi eI_{0}\delta(\omega-\omega^{\prime}) \ .
\label{eq:s4}
\end{equation}

\noindent The following representation of the delta function has been
used here:

\begin{equation}
\delta(\omega_{0}-\omega) =
\frac{\sin[(\omega_{0}-\omega)T}{\pi(\omega_{0}-\omega)} \qquad
{\mathrm{as}} \quad T \to \infty \ .
\label{eq:df1}
\end{equation}

\noindent The definition given in (\ref{eq:df1}) can be used to prove
the basic property of the delta function. Let $f(\omega)$ be any
function of $\omega$ that is non-singular at $\omega = \omega_{0}$, and
consider the integral

\begin{displaymath}
\int\limits^{\omega_{2}}_{\omega_{1}}f(\omega)\delta(\omega_{0}
-\omega)\D\omega = \frac{1}{\pi}\lim_{T\to\infty}
\int\limits^{\omega_{2}}_{\omega_{1}}f(\omega)\frac{\sin[(\omega_{0}
-\omega)T]}{(\omega_{0}-\omega)}\D\omega \ .
\end{displaymath}

\noindent Changing the variable of integration to $x =
(\omega-\omega_{0})T$ gives

\begin{eqnarray}
& \mbox{} &
\int\limits^{\omega_{2}}_{\omega_{1}}f(\omega)\delta(\omega_{0}
-\omega)\D\omega = \frac{1}{\pi}\lim_{T\to\infty}
\int\limits^{(\omega_{2}-\omega_{0})T}
_{(\omega_{1}-\omega_{0})T}
f((x/T)+\omega_{0})\frac{\sin x}{x}\D x
\nonumber\\
& \mbox{} &
=
\frac{1}{\pi}f(\omega_{0})
\int\limits^{\infty}_{-\infty}
\frac{\sin x}{x}\D x = f(\omega_{0}) \ ,
\quad {\mathrm{proved}}\quad \omega_{1} < \omega_{0}
< \omega_{2} \ .
\label{eq:df2}
\end{eqnarray}

\noindent Inclusion of the delta function in an integral therefore picks
out the value of the integrand at the point specified by the constant
in the delta function.

Thus, if a current of mean value $I_{0}$ flows in the input
of an amplifier whose frequency response curve is $G(\omega)$, the
output fluctuations are

\begin{equation}
\langle (\delta I_{\mathrm{out}})^{2}\rangle =
2\langle\mid\tilde{I}(t)\mid^{2}\rangle =
\frac{eI_{0}}{\pi}\int\limits^{\infty}_{0}\mid
G(\omega)\mid^{2}\D\omega \ .
\label{eq:s5}
\end{equation}

\noindent We can also write (\ref{eq:s5}) in the form

\begin{equation}
\langle (\delta I_{\mathrm{out}})^{2}\rangle =
\frac{eI_{0}}{\pi}\mid G(\omega_{0})\mid^{2}
\Delta\omega_{\mathrm{A}} \ ,
\label{eq:s6}
\end{equation}

\noindent where the effective amplification bandwidth of the klystron
is defined as

\begin{displaymath}
\Delta\omega_{\mathrm{A}} = \int\limits^{\infty}_{0}
\frac{\mid G(\omega)\mid^{2}}{\mid G(\omega_{0})\mid^{2}}\D\omega \ .
\end{displaymath}

\noindent We can further express this result by saying that the mean
square value of the output is equivalent to a fluctuating current of
mean square value

\begin{displaymath}
\langle(\delta I_{\mathrm{in}})^{2}\rangle
= eI_{0}\Delta\omega_{\mathrm{A}}/(\pi)
\end{displaymath}

\noindent at the input. Note that it has not been necessary to make any
assumptions about the amplification process. The result is completely
general; that is, it applies for any type of linear vacuum-tube
amplifier \cite{n1}.

With the preceding results in hand, it should now be possible to
estimate, for example, the amplitude of initial beam density modulation
$a_{\mathrm{in}}$. For this purpose it is convenient to express the
result (\ref{eq:s5}) in a different form. Noting that $\langle(\delta
I_{\mathrm{in}})^{2}\rangle = \langle a^{2}_{\mathrm{in}}\rangle
I^{2}_{0}$ and $\Delta\omega_{\mathrm{A}}/\omega_{0} \simeq
N^{-1}_{\mathrm{w}}$ we can write

\begin{equation}
\langle a^{2}_{\mathrm{in}}\rangle \simeq
e\omega_{0}/(I_{0}\pi N_{\mathrm{w}}) =
1/N_{\mathrm{c}} \ ,
\label{eq:s10}
\end{equation}

\noindent where $N_{\mathrm{c}} = N_{\lambda}N_{\mathrm{w}}/2$ is
the number of cooperating electrons, and $N_{\lambda} = 2\pi
I_{0}/(e\omega_{0})$ is the number of electrons per radiation
wavelength. If $I_{0} = 100 \ {\mathrm{A}}$, $\lambda = 30 \
{\mathrm{nm}}$, and $N_{\mathrm{w}} = 100$, appropriate substitutions
in (\ref{eq:s10}) show that the mean square value of input
is about $\langle a^{2}_{\mathrm{in}}\rangle \simeq
3\cdot 10^{-7}$.

Now we return to the question about the electron bunch profile.
Equation (\ref{eq:s3}) is not only
true at the arbitrary chosen rectangular bunch profile, but of course
it is true at any other bunch profile, and thus under the condition
(\ref{eq:s8}) in the long electron bunch limit the output intensity
fluctuations are all the same. (The final statement (\ref{eq:s5}) does
not involve the bunch profile function, which appeared only in the
intermediate arguments.)  Let us discuss the region of validity of the
approximation (\ref{eq:s8}) for different bunch profile functions.  If
we consider the case of rectangular bunch profile, we find that the
region of applicability of condition (\ref{eq:s8}) is less than that
for the case of a Gaussian bunch of the same duration. This is due to
the fact that the bunch form factor $\mid\bar{F}(\omega)\mid^{2}$,
decreas more slowly with the increase in frequency. On the other hand,
in a realistic situation there is no sharp boundary of the bunch and
the beam current falls to zero during some time interval $\Delta
\sigma_{\mathrm{T}} \ll T$. When the beam current at the edge falls in
accordance with a Gaussian law, $\Delta\sigma_{\mathrm{T}}$ must obey
the following conditions:

\begin{displaymath}
\Delta\sigma_{\mathrm{T}}/T \ll 1 \ , \quad
\Delta\sigma_{\mathrm{T}}\omega \gg 1 \ ,
\quad N/(\Delta\sigma_{\mathrm{T}}\omega)^{4}(\omega T)^{2} \ll 1 \ .
\end{displaymath}

\noindent When $\omega T \simeq 10^{4}$ and
$\omega\Delta\sigma_{\mathrm{T}} \simeq 10^{2}$ the value
$(\Delta\sigma_{\mathrm{T}}\omega)^{4}(\omega T)^{2}$ is equal to
$10^{16}$. As a rule, the
number of particles in the bunch, $N$, is less than or about $10^{11}$,
so condition (\ref{eq:s8}) is fulfilled.

\subsection {Temporal coherence}

The shot noise in the electron beam is a Gaussian random process.
The klystron amplifier operated in linear regime can be considered as
a linear filter which does not change the statistics of the signal. As
a result, we can define general properties of the beam density
modulation after the dispersion section without any calculations. For
instance, the square of instantaneous amplitude of density modulation
fluctuates in accordance with the negative exponential distribution. We
can also state that the spectral density of the electron beam
modulation and first-order time correlation function should form a
Fourier transform pair (this is the Wiener Khintchine theorem).

The correlation between the output currents at times $t$ and
$t^{\prime}$ has the form:

\begin{displaymath}
\langle\delta I_{\mathrm{out}}(t)\delta
I_{\mathrm{out}}(t^{\prime})\rangle =
\langle\tilde{I}_{\mathrm{out}}(t)\tilde{I}^{*}_{\mathrm{out}}
(t^{\prime})\rangle
\exp[-\I\omega_{0}(t-t^{\prime})]  + {\mathrm{C.C.}}
\end{displaymath}

\noindent Using (\ref{eq:s7}) and (\ref{eq:s4}) we calculate the
correlation between the complex amplitude $\tilde{I}(t)$ and $\tilde
{I}^{*}(t^{\prime})$:

\begin{eqnarray}
& \mbox{} &
\langle\tilde{I}_{\mathrm{out}}(t)\tilde{I}^{*}_{\mathrm{out}}
(t^{\prime})\rangle
\exp[-\I\omega_{0}(t-t^{\prime})]
\nonumber\\
& \mbox{} &
=
\frac{1}{4\pi^{2}}\int\limits^{\infty}_{0}\D\omega
\int\limits^{\infty}_{0}\D\omega^{\prime}
\exp(-\I\omega t + \I\omega^{\prime}t^{\prime})
G(\omega)G^{*}(\omega^{\prime})
\langle\bar{I}(\omega)\bar{I}^{*}(\omega^{\prime})\rangle
\nonumber\\
& \mbox{} &
=
\frac{eI_{0}}{2\pi}
\int\limits^{\infty}_{0}\D\omega
\exp[-\I\omega(t-t^{\prime})]\mid G(\omega)\mid^{2} \ .
\label{eq:cf}
\end{eqnarray}

\noindent We define the first-order time  correlation function as
follows:

\begin{displaymath}
g_{1}(t-t^{\prime}) = \frac{\langle\tilde{I}(t)\tilde{I}^{*}
(t^{\prime})\rangle}{\sqrt{\langle\mid\tilde{I}(t)\mid^{2}\rangle
\langle\mid\tilde{I}(t^{\prime})\mid^{2}\rangle}}
\end{displaymath}

\noindent Using (\ref{eq:d1}), (\ref{eq:s5}) and (\ref{eq:cf}) we can
write

\begin{displaymath}
g_{1}(t-t^{\prime}) =   \frac{\int\limits^{\infty}_{0}\D\omega
\mid G(\omega)\mid^{2}\exp[-\I(\omega-\omega_{0})(t-t^{\prime})]}
{\int\limits^{\infty}_{0}\D\omega\mid G(\omega)\mid^{2}} \ .
\end{displaymath}

\noindent The integral should go from $0$ to $\infty$, but $0$ is so
far from $\omega_{0}$ that the curve is all
finished by that time , so we go instead to minus $\infty$ - it makes
no difference. If we do that we get

\begin{equation}
g_{1}(t-t^{\prime}) =
\frac{\int\limits^{\infty}_{-\infty}\D (\Delta\omega) \mid
G(\Delta\omega)\mid^{2}\exp[-\I(\Delta\omega)(t-t^{\prime})]}
{\int\limits^{\infty}_{-\infty}\D(\Delta\omega)\mid
G(\Delta\omega)\mid^{2}} \ .
\label{eq:cf1}
\end{equation}

\noindent where $\Delta\omega = (\omega-\omega_{0})$. Therefore, the
slowly varying correlation function and the normalized spectrum of the
narrow band-signal are a Fourier transform pair. It is seen from latter
expression that the first-order correlation function possess the
property $g_{1}(t-t^{\prime}) = g_{1}^{*}(t^{\prime}-t)$. When the
klystron gain curve is symmetrical with respect to the resonance
frequency, $\omega_{0}$, the function $g_{1}$ is real. In our
approximation, the klystron gain function $\mid
G(\Delta\omega)\mid^{2}$ is symmetric with respect to $\Delta\omega$
and the integration therefore needs to be performed only from
$0$ to $\infty$ with the result being doubled. Inserting (\ref{eq:df})
into the integrand (\ref{eq:cf1})  we get

\begin{displaymath}
g_{1}(\tau) = \frac{\int\limits^{\infty}_{0}
f(x)\cos(x\tau/\alpha)\D x}{\int\limits^{\infty}_{0}f(x)\D x} \ .
\end{displaymath}

\begin{figure}[tb]
\begin{center}
\epsfig{file=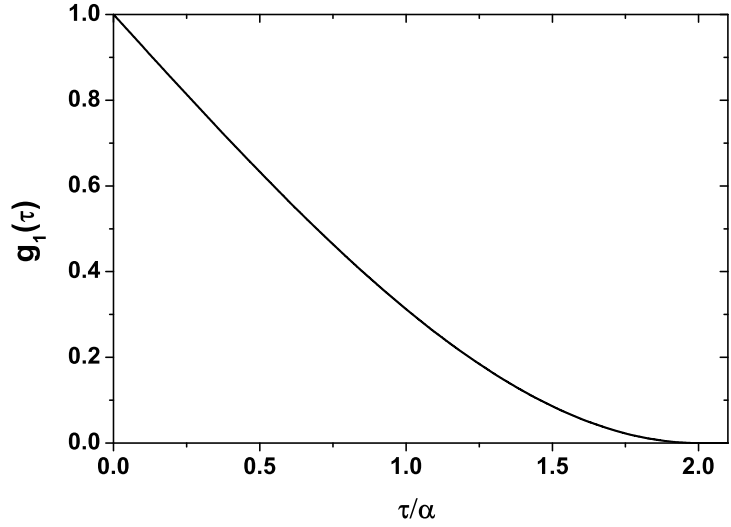,width=0.8\textwidth}
\end{center}
\caption{First-order time correlation function}
\label{fig:g1}
\end{figure}

\noindent Here we use the following notation $\alpha = \pi
N_{\mathrm{w}}/\omega_{0}$, $\ \tau = t-t^{\prime}$. The integral is a
Fourier transform of $f(x)$.
In Fig. \ref{fig:g1} we present a plot of first-order time
correlation function. First, let us
notice a remarkable feature of that plot. The correlation
function is zero for $\mid\tau\mid \ > 2\alpha =
N_{\mathrm{w}}\lambda/c$.  Of course we could predict such a result.
Remembering that the synchronism takes place when the electromagnetic
wave advance the electron beam by one wavelength at one undulator
period (see section 3), we see that $2\alpha$ is the total relative
slippage of the wave with respect to the electron beam.

In many applications it is desirable to have a precise and
definite meaning for the term "coherence time." Such a definition can
be made in terms of the complex correlation function, but there are a
multitude of definitions in terms of $g_{1}(t-t^{\prime})$ that can be
imagined.  However, in future discussions there is one definition that
arises most naturally and most frequently. Following the approach of
Mandel, we define the coherence time, $\tau_{\mathrm{c}}$, as

\begin{displaymath}
\tau_{\mathrm{c}} = \int\limits^{\infty}_{-\infty}\mid
g_{1}(\tau)\mid^{2}\D\tau \ .
\end{displaymath}

\begin{figure}[tb]
\begin{center}
\epsfig{file=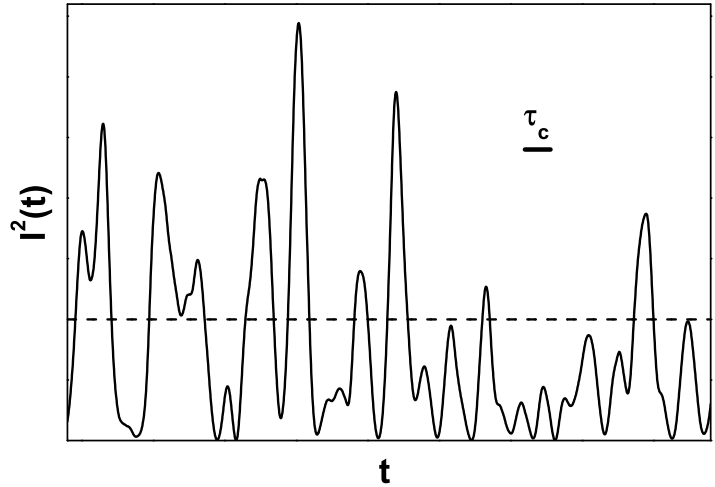,width=0.8\textwidth}
\end{center}
\caption{Time dependence of the cycle-averaged output intensity
modulation corresponds
to the shot noise fluctuations in the input beam current, obtained
from a computer simulation.  The dashed line shows the mean square
value of the output current fluctuations averaged over a time long
compared to $\tau_{\mathrm{c}}$ }
\label{fig:fl}
\end{figure}


Its magnitude is of the order of the inverse of the output frequency
spread. In all theory that follows, attention is restricted
to modulated electron beams whose frequency spreads are small compared
with the mean frequency, that is, where $\omega_{0}\tau_{\mathrm{c}}$
is very much larger than unity. Figure \ref{fig:fl}  illustrates the
type of fluctuations that occur in the cycle-averaged intensity
modulation. The figure has been constructed by a computer simulation
of the first cascade output in which the summation is carried out
explicitly for the real number of electrons in the electron bunch.  The
instantaneous intensity modulation varies in an unpredictable manner
about the average value, and it is this kind of variation that
characterizes the phenomenon as noise.  It is seen that substantial
changes in the square value of amplitude occur over a time span
$\tau_{\mathrm{c}}$, but this quantity is reasonably constant over
time intervals $\Delta t \ll \tau_{\mathrm{c}}$. The value of the
coherence time for the case of klystron is given by $\tau_{\mathrm{c}}
= 0.94\alpha = 0.47N_{\mathrm{w}}\lambda/c$. With the numerical values
$N_{\mathrm{w}} = 100$ and $\lambda = 30 \
{\mathrm{nm}}$ the value of coherence time is $\tau_{\mathrm{c}}
\simeq 5 \ {\mathrm{fs}}$.

Our discussion of the current fluctuations would be incomplete if we
did not refer the reader to another method of average calculation.
The klystron output coherence time, $\tau_{\mathrm{c}}$, is in
femtosecond range and $I_{0} =
{\mathrm{const.}}$ over the time span $\tau_{\mathrm{c}}$ is
a good assumption in all practical problems.  Clearly we can assume
that the current in question is adequately modelled as a ergodic and
hence stationary random process.  A stationary process obeys
statistical laws and is subject to conditions that do not vary with
time. For a stationary process time and ensemble averages are
equivalent.  In this case averaging symbol $\langle\cdots\rangle$
simply  means the average over a period much longer compare with
$\tau_{\mathrm{c}}$.

\section{Further applications of klystron}

In the preceding sections we have given a simple discussion of a soft
X-ray cascade klystron amplifier. We have not only neglected many
features of the klystron amplifier, but we have also completely omitted
some important applications. We shall mention three here: the frequency
multiplier, a scheme for femtosecond experiments, and multi-user soft
X-ray facility.

\subsection{Klystron frequency multipliers}

The remaining application of the klystron which we shall mention is the
frequency multiplier. We have seen, in section 3, that the bunched beam
at large values of the bunching parameter has not only a fundamental
radiation frequency component, but also has considerable intensity in
its harmonics. It is then possible to have an input undulator operating
at one frequency, and an output undulator operating at a multiple of
this frequency. The radiation in the output undulator will then be
excited by the harmonic component in the electron beam, and the
klystron will operate as a combination of frequency multiplier and
amplifier. A possible application would be to the production of very
high frequencies. A klystron gets more and more difficult to construct,
as the frequency becomes higher and higher, on account of the reduced
period of the undulators. If we use a frequency multiplier klystron to
reach the highest frequencies, it is only the output undulator which
must be made with the smallest period; the cascade undulators,
operating at a lower frequency, can be made with larger period, and
this can accommodate an injector with larger emittance and make the
source easier to construct.

The idea of using two undulators, with the second undulator resonant to
one of the harmonics of the first one (the so-called "after-burner"
scheme), was considered in \cite{b3,b1,b2}.  The technique of using an
conventional FEL amplifier as a frequency multiplier relies on the fact
that spatial bunching in the exponential gain regime can be very
strong.  The first uniform undulator is long enough to reach saturation
and produce strong spatial bunching in harmonics. The bunched beam
generates coherent radiation in the second undulator which
immediately follows the first one. The main problem with this approach
is the large induced (correlated) energy spread which will be generated
by the bunching of the electron beam at the fundamental frequency.
While this energy spread is necessary for the bunching, it degrades the
performance of the radiation section at the harmonic frequency.

The problem to be solved is how to prolong the interaction of the
bunched electron beam with the electromagnetic wave in the harmonic
radiator undulator. The present section is concerned with frequency
doubler, designed for conditions approximating maximum efficiency.
A reliable method to increase the frequency doubler
efficiency is to use a cascade klystron in which the input
undulator operates at the fundamental resonance, the penultimate
undulator operates at the fundamental harmonic, and the output
undulator operates at the second harmonic. In particular it can be
shown that a so-called 1-1-2 SASE scheme may operate with rather high
efficiency in the soft X-ray wavelength range. (The numbers 1,2 here
designate the resonant harmonic number for each undulator.)
The klystron gain is controlled in such a way that the maximum
energy modulation of the electron beam at the penultimate undulator
exit is about equal to the local energy spread. This is much smaller
than the induced energy spread of the electrons at the exit of
a saturated conventional SASE FEL. As a result, the operation of
a frequency doubler is based on prolonged interaction of the bunched
electron beam with the electromagnetic wave in the harmonic radiator.
In contrast to the after-burner method, in a two-cascade 1-1-2
scheme frequency doubler operation is based on prolonged interaction
of the bunched electron beam with the electromagnetic wave in the
output undulator and copious coherent emission at double frequency is
produced. The doubler efficiency can be optimized by monitoring the
output radiation intensity at double frequency as a function of the
compaction factor of the first and second dispersion section. In
\cite{fd} we have described an effective frequency doubler for SASE
FELs. For the first time the frequency multiplication scheme has been
analyzed for SASE FELs. Computer modeling  with time-dependent code
FAST  has demonstrated that the  final output power of a 1-1-2 klystron
doubler is no smaller than that which could be produced by a 2-2-2
klystron with the same electron beam parameters.

\subsection{Scheme for femtosecond experiments}

Time-resolved experiments are used to monitor time-dependent phenomena.
The standard technique for high-resolution time-resolved experiments is
the pump-probe scheme in which a process is started by a short pulse of
radiation (pump) and the evolution of the process is then observed
(probed) after a delay by means of a second
pulse of radiation, generally at another photon energy.
The obvious temporal limitation of the visible pump/X-ray probe
technique is the duration of the X-ray probe. Here we will concentrate
on the performance which can be obtained by a klystron XFEL operating
without bunch compression in the injector linac. At these sources, the
X-ray pulse duration is about 10 ps. This is longer than the timescale
of many interesting physical phenomena. The new
principle of pump-probe techniques described below offers a way around
this difficulty.

\begin{figure}[tb]
\begin{center}
\epsfig{file=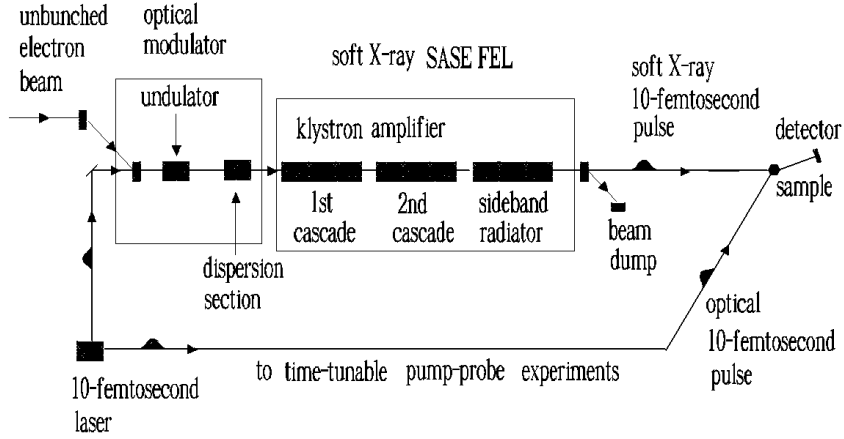,width=0.8\textwidth}
\end{center}
\caption{Scheme for time-resolved experiments based on the generation
of femtosecond pulses by sideband-seeded soft X-ray cascade
klystron amplifier. In this scheme, a femtosecond soft X-ray pulse is
naturally synchronized with the femtosecond optical pulse from the seed
laser and cancels jitter}
\label{fig:app}
\end{figure}

Our studies have shown that the soft X-ray cascade klystron
holds great promise as a source of radiation for generating high
power single femtosecond pulses. Our femtosecond soft X-ray
facility concept is based on the use of an X-ray SASE FEL combined with
a femtosecond quantum laser \cite{fs}. The operation of a femtosecond
soft X-ray SASE FEL is illustrated in Fig.  \ref{fig:app}. An
ultrashort laser pulse is used to modulate the density of electrons
within the femtosecond slice of the electron bunch at a frequency
$\omega_{\mathrm{opt}}$. The seed laser pulse will be timed to overlap
with the central area of the electron bunch. This ultrashort laser
pulse serves as a seed for a modulator which consists of an uniform
undulator and a dispersion section. The interaction
of seed pulse with the electron beam produces an energy modulation at
$\omega_{\mathrm{opt}}$.  This energy modulation is converted into
spatial bunching in the dispersion section. Density modulation at the
modulator exit is about 20 \%. The energy modulation introduced by the
modulator is smaller than the local energy spread.  Following the
modulator the beam enters the first undulator of the klystron amplifier
which is resonant with soft X-ray radiation at frequency $\omega_{0}$.
The process of amplification in the soft X-ray klystron amplifier
develops in the same way as was explained in previous sections:
fluctuations of the electron beam current density serve as the input
signal.  At the chosen level of density (energy) modulation the SASE
process develops nearly in the same way as with a non-modulated
electron beam because of the large ratio of cooperation length to
optical wavelength. By the time the beam is bunched in the klystron
cascades, at frequency $\omega_{0}$, the density
modulation amplitude has reached saturation. This leads to amplitude
modulation of the density at the sidebands
$\omega_{0}\pm\omega_{\mathrm{opt}}$. The sideband density modulation
takes place at the part of the electron pulse defined by the duration
of the seed laser pulse, which is much shorter than the electron pulse.
Following the last cascade the beam enters the output undulator which
is resonant at the frequency $\omega_{0} - \omega_{\mathrm{opt}}$.
Because the beam has a large component of bunching at the sideband,
coherent emission is copiously produced within the femtosecond slice of
the electron bunch. Analytical methods are of limited use in to the
study of the sideband-seeded SASE FEL and numerical simulations must be
used.  Simulations with time-dependent code FAST provide a "full
physics" description of the process  in a sideband-seeded SASE FEL. The
results of numerical simulations confirm our simple physical
considerations \cite{fs}.

We assume that the SASE
bandwidth is much less than the separation of the sidebands from the
main peak. This requirement is of critical importance to the overall
performance of the femtosecond SASE FEL facility. In this case,
the (output) undulator can be used to distinguish the sideband density
modulation from the density modulation at the central frequency.
Obviously, this requirement is easier to achieve for long radiation
wavelengths. For  300 nm laser radiation and for 30 nm output
radiation, for example, the amplification bandwidth of the cascade
klystron must be much smaller than 10 \%.  This condition may be easily
satisfied in practice.  The long (unbunched) electron pulse is one of
the advantages of the adopted FEL design. Since the electron pulse
duration (about 10 ps) is much longer than the time jitter of the
electron and seed laser pulses $( {\mathrm{about}} \ \pm 1 \
{\mathrm{ps}})$, the synchronization of the optical
laser with the electron pulses is not a problem.

\begin{figure}[tb]
\begin{center}
\epsfig{file=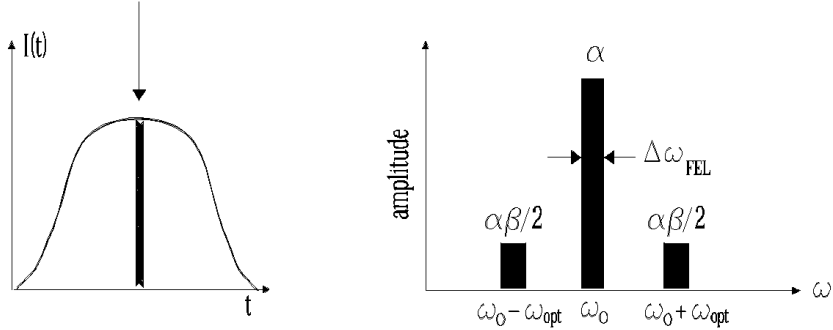,width=0.8\textwidth}
\end{center}
\caption{Description of the sideband generation for the case of the
density modulation as initial condition}
\label{fig:asb}
\end{figure}

Pump-probe techniques which are commonly used with optical lasers, are
highly desirable in order to make full use of the femtosecond soft
X-ray pulses. Since, in this case, precise timing is needed with a
jitter of less than 10 fs, in \cite{fs} we suggested to combine the
femtosecond soft X-ray pulses generated by the sideband-seeded SASE FEL
with optical pulses generated by the seed laser system. It should be
emphasized that in the proposed scheme a femtosecond soft X-ray pulse
is naturally synchronized with his femtosecond optical pulse and
cancels jitter. This development of XFEL based pump-probe experiments
allow us to investigate phenomena at timescale down to 100 fs.

\subsection{Multi-user distribution system for
XFEL laboratory}

An X-ray laboratory should serve several, may be up to ten
experimental stations which can be operated independently according to
the needs of the user community. On the other hand,
the multi-user distribution system has to satisfy
an additional requirement. Passing the electron bunch through the
bending magnets must avoid emittance dilution due to coherent
synchrotron radiation (CSR) effects. For very short bunches and very
high peak current, CSR can generated energy spread in the bending
magnets and thus dilute the horizontal emittance.  As a result, the
prefered layout of a conventional SASE FEL is a linear arrangement
in which the injector , accelerator, bunch compressors and undulators
are nearly collinear, and in which the electron beam does not change
direction between accelerator and undulators.

\begin{figure}[tb]
\begin{center}
\epsfig{file=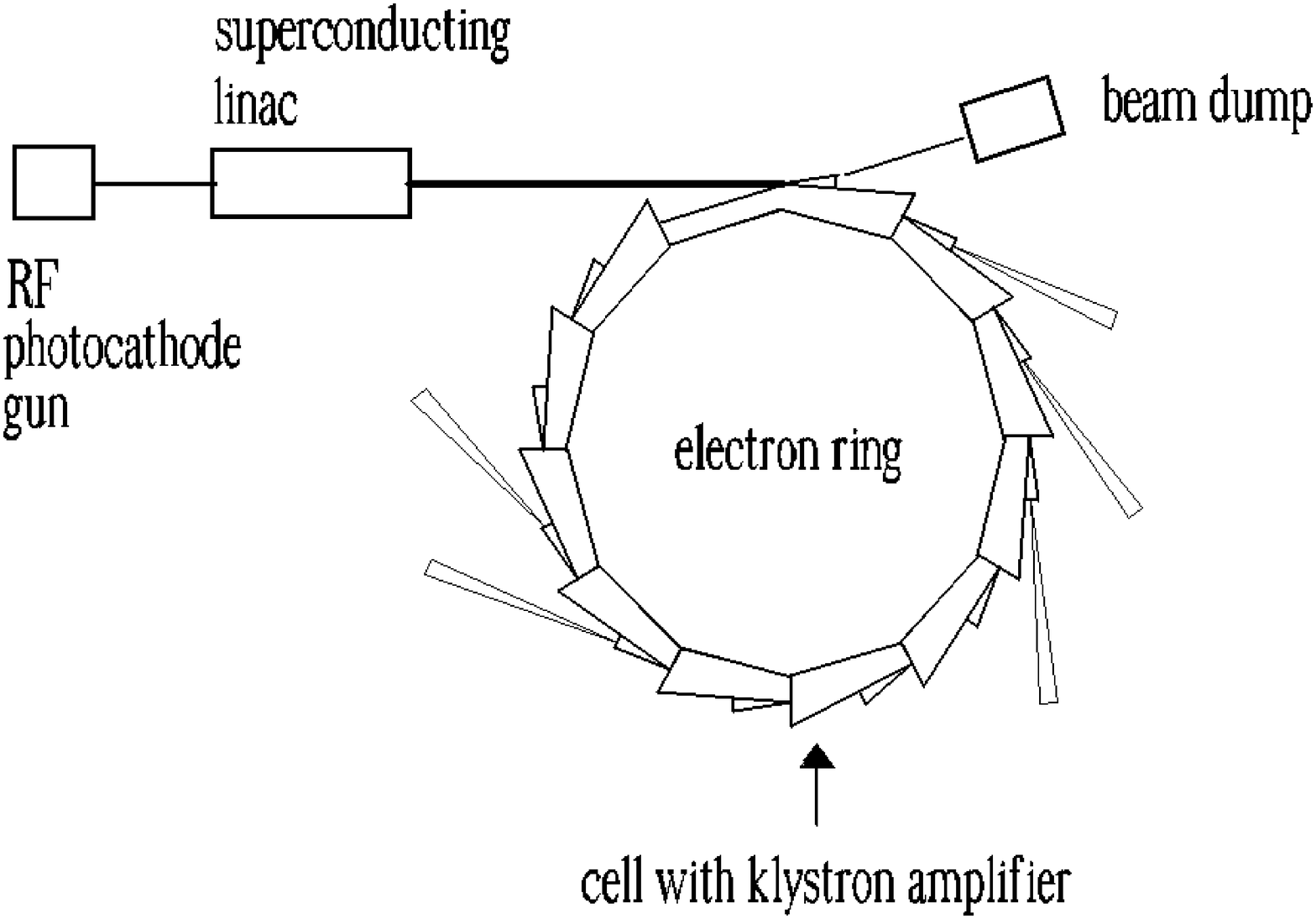,width=0.8\textwidth}
\end{center}
\caption{Diagram of a possible fourth-generation synchrotron facility
using spontaneous emission free-electron laser klystron amplifiers }
\label{fig:a4gf}
\end{figure}

\begin{figure}[tb]
\begin{center}
\epsfig{file=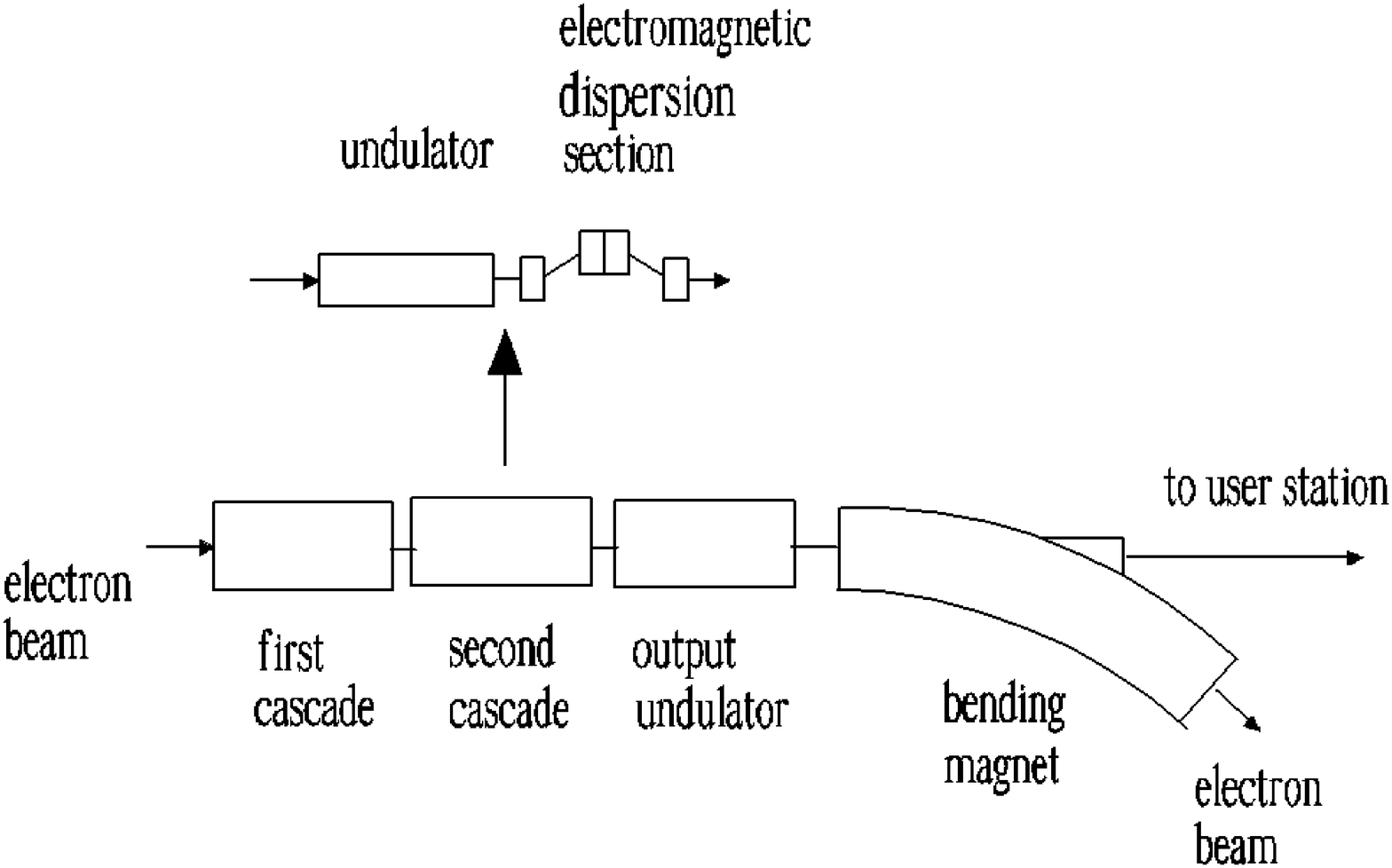,width=0.8\textwidth}
\end{center}
\caption{Electron ring cell design. Using a klystron
electromagnetic dispersion section as a switching element it is
possible to quickly switch off (on) the cell klystron amplifier thus
providing multi-user capability.  This design makes it possible to make
various wavelengths of SASE radiation available in the XFEL laboratory
quasi-simultaneously}
\label{fig:acu}
\end{figure}

The situation
is quite different for the klystron amplifier scheme proposed in our
paper.  Since it operates without bunch compression in the injector
linac, the problem of emittance dilution in the bending magnets does
not exist. An electron beam distribution system based on
unbunched electron beam can provide efficient ways to generate a
multi-user facility - very similar to present day synchrotron radiation
facilities.  A possible layout of a soft X-ray FEL laboratory based on
an electron ring distribution system is shown in Fig. \ref{fig:a4gf}.
The layout of the laboratory follows a similar approach as that used
for synchrotron light sources. The SASE FEL user facility consists of
an injector linac and electron beam distribution system. The
injector is composed of a RF gun with photocathode and a main
(superconducting) linac. In order to make efficient use of the new
source it is proposed to segment the full circumference of a
distribution system into arcs which are repeated a number of times to
form a complete ring. Each cell includes a two-cascade klystron and a
bending magnet. A specific realization of the electron ring cell is
sketched in Fig.  \ref{fig:acu}. The electron beam transport line
guiding electrons from the injector linac to the experimental hall is
connected tangentially to one of the straight sections of the electron
ring. In order to obtain a useful separation between the experimental
areas behind the photon beam lines, an angle of 36 degrees between two
neighboring lines would be desirable. Thus, ten beam lines can be
installed on a complete electron ring. Using klystron (electromagnetic)
dispersion sections in each cell  as switching elements it is possible
to quickly switch the FEL photon beam from one experiment to the other,
thus providing multi-user capability. Users can define the radiation
wavelength for their experiment independently of each other to a very
large extent, since they use different undulators. Injector linac and
electron beam transport lines operate at fixed parameters. At a fixed
electron energy the magnet gap of the klystron undulators can be varied
mechanically for wavelength tuning.  This design makes it possible to
make various wavelengths of SASE radiation available in the XFEL
laboratory quasi-simultaneously.  It is a great advantage that injector
and electron beam transport lines in the new scheme of multi-user
facility operate at fixed parameters and that an "electron switchard"
is not required.

When a relativistic electron beam passes through the undulator, it
emits incoherent radiation. This process leads to an increase of
the energy spread in the beam due to quantum fluctuations of the
undulator radiation.  This effect grows significantly with an
increase in electron beam energy, strength field and length of
the undulator. This effect should be carefully taken into account when
designing multi-user distribution system. For the case of planar
undulator the expression for the rate of energy diffusion can be
written in the following form\cite{cf}:

\begin{equation}
\frac{\D\langle(\Delta\gamma)^{2}\rangle}{\D z} =
\frac{7}{15}\frac{\lambda_{\mathrm{c}}}{2\pi}r_{\mathrm{e}}\gamma^{4}
k^{3}_{\mathrm{w}}K^{2}F(K) \ ,
\label{eq:cfz}
\end{equation}

\noindent where $\lambda_{\mathrm{c}}$ is the Compton wavelength,
$r_{\mathrm{e}}$ is the classical electron radius, $F(K) \simeq 1.2
K$ for $K \gg 1$. The next numerical example illustrates the amplitude
of the effect. If $\gamma = 10^{3}$, $K = 1.42$, $\lambda_{\mathrm{w}}
= 3 \ {\mathrm{cm}}$, and the total undulator length is equal to 100 m,
appropriate substitution in (\ref{eq:cfz}) shows that the energy spread
increase due to quantum fluctuations is about
$\sqrt{\langle(\Delta{\cal E})^{2}\rangle} \simeq 1 \ {\mathrm{keV}}$
at the end of the 10th klystron amplifier, and has a negligible effect
on the klystron performance.

\section{Conclusion}

The high-gain klystron amplifier described in this paper is an
attractive alternative to other FEL configurations for
operation in short wavelength range. An distinguishing feature of the
klystron amplifier is the absence of apparent limitations which
would prevent operation without bunch compression in the injector
linac. We have illustrated the proposed cascade
klystron scheme using the parameters of the TESLA Test Facility.
Although the present work is concerned primarily for use in the soft
X-ray spectrum, its applicability is not restricted to that range,
for example an X-ray SASE FEL is a suitable candidate for application
of cascade klystron scheme described here.

\section*{Acknowledgments}

We thank R. Brinkmann, B. Faatz, J. Feldhaus, G. Geloni, M. Koerfer, J.
Pflueger, J. Rossbach for many useful discussions. We thank
C. Pagani, J.R.~Schneider and D.~Trines for interest in this work.

\end{document}